\documentclass[12pt,preprint]{aastex}
\usepackage[section] {placeins}
\shorttitle{CARMA OBSERVATIONS OF ORION-KL I}
\shortauthors{Widicus Weaver \& Friedel}
\begin{document}
\newcommand\acetone{(CH$_3$)$_2$CO}
\newcommand\dme{(CH$_3$)$_2$O}
\newcommand\mef{HCOOCH$_3$}
\newcommand\fa{HCOOH}
\newcommand\fmal{H$_2$CO}
\newcommand\mtoh{CH$_3$OH}
\newcommand\kms{km s$^{-1}$}
\newcommand\jbm{Jy/beam}
\newcommand\vycn{C$_2$H$_3$CN}
\newcommand\mtcn{CH$_3$CN}
\newcommand\etcn{C$_2$H$_5$CN}
\newcommand\etoh{C$_2$H$_5$OH}
\newcommand\acal{CH$_3$CHO}
\newcommand\vlsr{$v_{\rm LSR}$}

\title{COMPLEX ORGANIC MOLECULES AT HIGH SPATIAL RESOLUTION TOWARD ORION-KL I: SPATIAL SCALES}

\author{Susanna L. Widicus Weaver\altaffilmark{1} and Douglas N. Friedel\altaffilmark{2}  }

\altaffiltext{1}{Department of Chemistry, Emory University, Atlanta, GA  30322\\
email: susanna.widicus.weaver@emory.edu}
\altaffiltext{2}{Department of Astronomy, 1002 W. Green St., University of
Illinois, Urbana IL 61801\\
email: friedel@astro.illinois.edu}

\begin{abstract}
Here we present high spatial resolution ($<$1 arcsecond) observations of molecular emission in Orion-KL conducted using the Combined Array for Research in Millimeter-Wave Astronomy (CARMA).  This work was motivated by recent millimeter continuum imaging studies of this region conducted at a similarly high spatial resolution, which revealed that the bulk of the emission arises from numerous compact sources, rather than the larger-scale extended structures typically associated with the Orion Hot Core and Compact Ridge.  Given that the spatial extent of molecular emission greatly affects the determination of molecular abundances, it is important to determine the true spatial scale for complex molecules in this region.   Additionally, it has recently been suggested that the relative spatial distributions of complex molecules in a source might give insight into the chemical mechanisms that drive complex chemistry in star-forming regions.  In order to begin to address these issues, this study seeks to determine the spatial distributions of ethyl cyanide [\etcn], dimethyl ether [\dme], methyl formate [\mef], formic acid [\fa], acetone [\acetone], SiO, methanol [\mtoh], and methyl cyanide [\mtcn] in Orion-KL at $\lambda$ = 3 mm. We find that for all observed molecules, the molecular emission arises from multiple components of the cloud that include a range of spatial scales and physical conditions.  Here we present the results of these observations and discuss the implications for studies of complex molecules in star-forming regions.

\end{abstract}


\section{Introduction}

The Orion-KL region is a high-mass star-forming region that has been extensively studied because of its rich chemical inventory.  A literature search reveals that over 50 microwave, millimeter, and submillimeter spectral line surveys have been reported for this source.  Previous molecular observations of Orion-KL have involved both spectral line surveys to determine the source composition, and imaging studies to explore the relative distributions of molecules in this complicated region. Imaging studies of molecular emission have indicated that the chemistry of Orion-KL displays so-called nitrogen-oxygen ``chemical differentiation," with the emission from nitrogen-bearing molecules such as methyl cyanide [\mtcn] and ethyl cyanide [\etcn] tracing the Orion Hot Core, and emission from oxygen-bearing molecules such as methyl formate [\mef] and dimethyl ether [\dme] tracing the Orion Compact Ridge.  The concept of nitrogen-oxygen chemical differentiation is based on the positions of the emissions peaks for these two classes of molecules being spatially-separated.  In the case of the Orion-KL source, this concept dates back to the the first reported millimeter line survey \citep{blake87}, where the spectral lines from these two classes of molecules were found to have different rest velocities and were therefore assumed to occupy different parcels of gas.  The follow-up imaging studies that ranged over the last two decades confirmed this differentiation through imaging of molecular emission at $>$1 arcsecond spatial resolution.  The concept of spatially distinct regions containing N-bearing and O-bearing molecules in Orion-KL was first called into question by \cite{friedel08}, who showed that the acetone molecule [\acetone] traces the gas at the intersection of the regions shown to contain spatially-distinct N-bearing and O-bearing molecules.  Furthermore, a recent $\lambda$=3 mm continuum imaging study at subarcsecond spatial resolution shows that Orion-KL is comprised of numerous, compact, bright continuum sources \citep{friedel11}.  The results of this study suggest that much of the molecular emission may actually trace these compact regions, rather than the more extended structures traditionally associated with the Hot Core and Compact Ridge \citep{friedel11}.

Observational studies reporting molecular column densities and/or abundances in Orion-KL generally rely upon a set of assumptions regarding the molecular emission in this region.  These assumptions are based on the previous imaging studies of molecular emission in this source, and include:  1. Oxygen-bearing molecules are located in the Compact Ridge, while nitrogen-bearing molecules are located in the Hot Core;  2. The emission regions are approximately spherical and $\sim$5\arcsec~ in diameter for both sources; 3. Molecules from these sources have rotational temperatures in the 150 - 200 K range.  If these assumptions are not valid at higher spatial resolutions, this would have significant implications for chemical modeling of star-forming regions.  Current state-of-the-art models assume that all of the molecules are co-spatial \citep{Garrod08,Laas}.  Furthermore, molecular abundances and temperatures are commonly used as benchmarks for comparison between model results and observations.  Determination of molecular abundances in Orion-KL relies upon an accurate source structure model for each molecule so that appropriate beam dilution corrections can be applied.  In addition to these important considerations, it has also recently been suggested that chemical imaging of interstellar sources can be used to probe the chemical mechanisms at work in these environments \citep{neill11}.  If various classes of complex molecules are shown to occupy different parcels of gas, hydrodynamic models will be required to explain this chemical differentiation and the underlying chemistry and physics that drives such a separation.  High-resolution imaging studies of the molecular distributions in sources like Orion-KL are therefore crucial for the advancement of chemical models of hot cores.

To this end, we have carried out interferometric observations over a range of spatial resolutions toward Orion-KL using the Combined Array for Research in Millimeter-Wave Astronomy (CARMA).  These spectral line and imaging observations target several complex molecules previously detected in this region, including acetone [\acetone], dimethyl ether [\dme], methyl formate [\mef], ethyl cyanide [\etcn], formic acid [\fa], and acetaldehyde [\acal].  We present here the spatial distributions and scales of the observed molecules and discuss the implications of these results in terms of the chemistry of the Orion-KL region. Additional information pertaining to the kinematics of Orion-KL based on these molecular observations are presented in a follow-up study (Friedel \& Widicus Weaver 2012, submitted; hereafter, Paper II).

\section{Observations}
CARMA was used in its B, D, and A configurations to conduct observations in 2007 December, 2008 July, and 2009 January, respectively. Two 7 hour tracks were taken in B configuration, one 6 hour track was taken in D configuration, and four 5-6 hour tracks in A configuration. The  phase center for all observations was $\alpha$(J2000) = $05^h35^m14^s.35$ and $\delta$(J2000) = $-05{\degr}22{\arcmin}35{\arcsec}.0$. The synthesized beam sizes are $\sim5.9\arcsec\times4.8\arcsec$ for the D configuration, $\sim1.1\arcsec\times0.9\arcsec$ for the B configuration, and $\sim0.4\times0.35\arcsec$ for the A configuration. One arcsecond is $\sim$414 AU at the distance of Orion-KL.  The projected baselines of the observations are 3.0-35.3 k$\lambda$ (10-118 m, D configuration), 21.8-257.5 k$\lambda$ (73-860 m, B configuration),  and 38.0-467.2 k$\lambda$ (127-1560 m, A configuration). Structures larger than $\sim$30 arcseconds are resolved out for observations with this range of $u-v$ coverage.

The correlator setup for the observations in the B and D configurations included six 31 MHz wide windows for observations of continuum and spectral lines (three in each sideband), with 63 channels per window, yielding a channel spacing of 488 kHz ($\sim$1.4 \kms). The flux density was calibrated using observations of Uranus and Mars, and the antenna-based gains were calibrated using observations of 0541-056. For two of the A configuration tracks, the correlator setup included four 31 MHz wide windows and two 500 MHz wide windows (three windows in each sideband). A third A configuration track was conducted with four 31 MHz windows and eight 500 MHz windows, and the fourth track with eight 31 MHz windows and four 62 MHz windows. Self-calibration on the SiO maser at 86.243 GHz was used to perform the antenna-based gain calibration for the A configuration tracks, with the solution being bootstrapped to the other windows. Observations of 0607-085 were used to determine and remove phase offsets between the other windows and the SiO window. The high spatial resolution in the A configuration observations led to the flux calibrators being heavily resolved; comparison to previously measured fluxes of 0607-085 allowed amplitude gains to be calculated for these A configuration observations, and the resultant amplitudes are accurate to within $\sim$20\%. The passband of the 31 MHz windows was corrected using the internal noise source.  The passbands of the 62 MHz and 500 MHz windows were corrected using observations of 0423-013. All calibration, continuum subtraction, and imaging was performed using the MIRIAD software package \citep{sault95}.

\section{Results and Discussion}
The transitions targeted in these observations are listed in Table~\ref{tab:mols}. This table includes the rest frequencies, quantum numbers, symmetry label, upper state energy, and linestrength information for each transition.  This table also includes information pertaining to the details of the observations, including beam size, RMS noise level, and array configuration for each observation.  Note that in many cases, lines arising from several of the transitions are blended.  Rotational temperatures cannot be reliably determined from such a small number of lines.  Therefore, the relative contributions of each transition to the overall flux of any given blended line also cannot be determined from this information.  The transitions listed in the lower section of the table (below the line) were detected in single channels of the wideband windows of the A configuration observations. While no velocity information can be determined for these transitions, they can be mapped to give spatial information.

\begin{deluxetable}{llrrrcrcc}
\tablecolumns{9}
\tabletypesize{\scriptsize}
\tablewidth{0pt}
\tablecaption{Molecular Parameters of Observed Lines}
\tablehead{\colhead{} & \colhead{Quantum} & \colhead{Frequency} &
\colhead{$E_u$} & \colhead{S$\mu^2$} &
\colhead{$\theta_a\times\theta_b$} & \colhead{RMS Noise} &
\colhead{Array} & \colhead{}\\
\colhead{Molecule} & \colhead{Numbers} & \colhead{(MHz)} & \colhead{(K)}
&\colhead{($D^2$)}& \colhead{($\arcsec\times\arcsec$)} &
\colhead{(m\jbm)} & \colhead{Config.} & \colhead{Ref.}}
\startdata
\etcn & $10_{2,9}-9_{2,8}$ & 89,297.660 (50) & 28 & 119.3 & $1.15\times0.93$ & 37.3 & B & 1\\
 & & & & & $6.1\times4.9$ & 20.5 & D & \\
 & & & & & $0.44\times0.34$ & 16.8 & A & \\
\mef & $8_{1,8}-7_{1,7} A$ & 89,314.589 (13) & 20 & 21.0 & $1.15\times0.93$ & 37.3 & B & 2\\
 & & & & & $6.1\times4.9$ & 20.5 & D & \\
 & & & & & $0.44\times0.34$ & 16.8 & A & \\
\mef & $8_{1,8}-7_{1,7} E$ & 89,316.668 (13) & 20 & 21.0 & $1.15\times0.93$ & 37.3 & B & 2\\
 & & & & & $6.1\times4.9$ & 20.5 & D & \\
 & & & & & $0.44\times0.34$ & 16.8 & A & \\
\dme & $15_{2,13}-15_{1,14}AE/EA$ & 88,706.220 (2) & 117 & 11.6 & $1.08\times0.88$ & 45.5 & B & 3\\
 & & & & & $6.1\times5.0$ & 21.1 & D & \\
\dme & $15_{2,13}-15_{1,14}EE$ & 88,707.701 (2) & 117 & 11.6 & $1.08\times0.88$ & 45.5 & B & 3\\
 & & & & & $6.1\times5.0$ & 21.1 & D & \\
\dme & $15_{2,13}-15_{1,14}AA$ & 88,709.181(3) & 117 & 11.6 & $1.08\times0.88$ & 45.5 & B & 3\\
 & & & & & $6.1\times5.0$ & 21.1 & D & \\
\acetone & $9_{2,8}-8_{1,7}AE$ & 101,427.041 (10) & 27 & 7.2 & $0.88\times0.82$ & 29.2 & B & 4\\
\acetone & $9_{2,8}-8_{1,7}EA$ & 101.427.130 (9) & 27 & 7.2 & $0.88\times0.82$ & 29.2 & B & 4\\
\acetone & $9_{2,8}-8_{1,7}EE$ & 101,451.059 (7) & 27 & 7.2 & $0.88\times0.82$ & 29.2 & B & 4\\
\acetone & $9_{2,8}-8_{1,7}AA$ & 101,475.733 (11) & 27 & 7.2 & $0.88\times0.82$ & 29.2 & B & 4\\
\acetone & $9_{1,8}-8_{2,7}AE$ & 101,426.664 (10) & 27 & 7.2 & $0.88\times0.82$ & 29.2 & B & 4\\
\acetone & $9_{1,8}-8_{2,7}EA$ & 101,426.759 (9) & 27 & 7.2 & $0.88\times0.82$ & 29.2 & B & 4\\
\acetone & $9_{1,8}-8_{2,7}EE$ & 101,451.446 (7) & 27 & 7.2 & $0.88\times0.82$ & 29.2 & B & 4\\
\acetone & $9_{1,8}-8_{2,7}AA$ & 101,475.332 (11) & 27 & 7.2 & $0.88\times0.82$ & 29.2 & B & 4\\
\acetone & $9_{0,9}-8_{1,8}AE$ & 92,727.904 (11) & 24 & 8.4 & $5.9\times4.8$ & 24.9 & D & 4\\
\acetone & $9_{0,9}-8_{1,8}EA$ & 92,727.950 (10) & 24 & 8.4 & $5.9\times4.8$ & 24.9 & D & 4\\
\acetone & $9_{0,9}-8_{1,8}EE$ & 92,735.670 (8) & 24 & 8.4 & $5.9\times4.8$ & 24.9 & D & 4\\
\acetone & $9_{0,9}-8_{1,8}AA$ & 92,743.361 (12) & 24 & 8.4 & $5.9\times4.8$ & 24.9 & D & 4\\
\acetone & $9_{1,9}-8_{0,8}AE$ & 92,727.907 (11) & 24 & 8.4 & $5.9\times4.8$ & 24.9 & D & 4\\
\acetone & $9_{1,9}-8_{0,8}EA$ & 92,727.953 (10) & 24 & 8.4 & $5.9\times4.8$ & 24.9 & D & 4\\
\acetone & $9_{1,9}-8_{0,8}EE$ & 92,735.673 (8) & 24 & 8.4 & $5.9\times4.8$ & 24.9 & D & 4\\
\acetone & $9_{1,9}-8_{0,8}AA$ & 92,743.364 (12) & 24 & 8.4 & $5.9\times4.8$ & 24.9 & D & 4\\
\fmal & $6_{1,5}-6_{1,6}$ & 101,332.991 (44) & 88 & 5.1 & $0.85\times0.80$ & 36.9 & B & 5\\
$^{34}$SO & $3_2-2_1$ & 97,715.390 (150) & 9 & 6.9 & $0.79\times0.78$ & 20.2 & B & 5\\
\acal & $5_{1,4}-4_{1,3}$ & 98,863.314 (6) & 17 & 30.4 & $0.87\times0.82$ & 36.7 & B & 6\\
\fa & $4_{1,4}-3_{1,3}$ & 86,546.185 (17) & 14 & 7.3 & $1.11\times0.90$ & 44.6 & B & 5\\
\fa & $4_{1,3}-3_{1,2}$ & 93,098.355 (17) & 14 & 7.3 & $5.9\times4.8$ & 21.3 & D & 5\\
SiO & 2-1 & 86,846.960 (50) & 6.2 & 19.3 & $0.49\times0.37$ & 42.8 & A & 5\\
\mtoh & 7$_{2,6}-6_{3,3}A-$ & 86,615.602 (14) & 103 & 1.4 & $0.49\times0.37$ & 15.5 & A & 7\\
\hline
\dme & $3_{2,2}-3_{2,1}EE$ & 91,476.596 (2) & 11.1 & 2.4 & $0.48\times0.37$ & 15.1 & A & 3\\
CH$_3$CN & $5_4-4_4$\tablenotemark{a} & 91,958.728 (36) & 127 & 27.7  & $0.49\times0.36$ & 2.3 & A & 8\\
CH$_3$CN & $5_3-4_3$\tablenotemark{a} & 91,971.132 (25) & 77 & 49.2  & $0.49\times0.36$ & 2.3 & A & 8\\
CH$_3$CN & $5_2-4_2$\tablenotemark{a} & 91,979.997 (17) & 42 & 64.5  & $0.49\times0.36$ & 2.3 & A & 8\\
CH$_3$CN & $5_1-4_1$\tablenotemark{a} & 91,985.317 (16) & 20 & 73.8  & $0.49\times0.36$ & 2.3 & A & 8\\
CH$_3$CN & $5_0-4_0$\tablenotemark{a} & 91,987.090 (16) & 13 & 76.8  & $0.49\times0.36$ & 2.3 & A & 8\\
CH$_3$CN & $5_{3,-1}-4_{3,-1}~^1\nu_8$\tablenotemark{b} & 92,234.587 (20) & 649 & 49.2 & $0.48\times0.36$ & 1.9 & A & 8\\
CH$_3$CN & $5_{2,-1}-4_{2,-1}~^1\nu_8$\tablenotemark{b} & 92,247.254 (16) & 601 & 64.5 & $0.48\times0.36$ & 1.9 & A & 8\\
CH$_3$CN & $5_{4,2}-4_{4,2}~^1\nu_8$\tablenotemark{b} & 92,249.629 (26) & 599 & 27.7 & $0.48\times0.36$ & 1.9 & A & 8\\
CH$_3$CN & $5_{1,1}-4_{1,1}~^1\nu_8$\tablenotemark{b} & 92,256.278 (200) & 559 & 73.8 & $0.48\times0.36$ & 1.9 & A & 8\\
CH$_3$CN & $5_{3,1}-4_{3,1}~^1\nu_8$\tablenotemark{b} & 92,258.402 (20) & 570 & 49.2 & $0.48\times0.36$ & 1.9 & A & 8\\
CH$_3$CN & $5_{0,1}-4_{0,1}~^1\nu_8$\tablenotemark{b} & 92,261.420 (160) & 546 & 76.8 & $0.48\times0.36$ & 1.9 & A & 8\\
CH$_3$CN & $5_{2,1}-4_{2,1}~^1\nu_8$\tablenotemark{b} & 92,263.938 (22) & 547 & 64.5 & $0.48\times0.36$ & 1.9 & A & 8\\
\vycn & $9_{5,5}-8_{5,4}$\tablenotemark{c} & 85,419.954 (10) & 75 & 90.5 & $0.52\times0.39$ & 1.4 & A & 1\\
\vycn & $9_{3,7}-8_{3,6}$\tablenotemark{c} & 85,426.926 (10) & 40 & 116.4 & $0.52\times0.39$ & 1.4 & A & 1\\
\vycn & $9_{3,6}-8_{3,5}$\tablenotemark{c} & 85,434.529 (10) & 40 & 116.4 & $0.52\times0.39$ & 1.4 & A & 1\\
\vycn & $9_{7,3}-8_{7,2}$\tablenotemark{c} & 85,447.694 (10) & 126 & 51.7 & $0.52\times0.39$ & 1.4 & A & 1\\
\vycn & $9_{8,2}-8_{8,1}$\tablenotemark{c} & 85,468.363 (10) & 158 & 27.4 & $0.52\times0.39$ & 1.4 & A & 1\\
\enddata
\tablenotetext{a}{These CH$_3$CN transitions are blended at our spectral resolution.}
\tablenotetext{b}{These CH$_3$CN$^1\nu_8$ transitions are blended at our spectral resolution.}
\tablenotetext{c}{These \vycn\ transitions are blended at our spectral resolution.}
\tablerefs{(1) \citet{brauer09}; (2)
\citet{oest99}; (3) \citet{gron98}; (4) \citet{gron02}; (5) \citet{jpl};
(6) \citet{lovas04}; (7) \citet{xu97}; (8) \citet{bou80}; (9) \citet{kisiel09}}
\label{tab:mols}
\end{deluxetable}

\clearpage

The spectral line and imaging results for the observations are presented in Sections \ref{sec:sio} - \ref{mtoh} below.  A description of continuum sources in the region was given in \citet{friedel11}, and many of the figures presented here reference sources labeled in Figure~2 of that work.  In addition, for the spectral line observations, a comparison of the flux observed in B ($\sim1\arcsec$ resolution) and D ($\sim5\arcsec$ resolution) configurations was required to determine whether any flux was resolved out in the higher spatial resolution observations.  To compute the B configuration flux for each source, the positive data were convolved with a Gaussian beam of the same size as the D configuration data, using the MIRIAD task CONVOL.  These results are summarized in Table \ref{tab:resolve} below.
\begin{deluxetable}{lrrr}
\tablecolumns{4}
\tabletypesize{\scriptsize}
\tablewidth{0pt}
\tablecaption{Extended Flux}
\tablehead{\colhead{} & \multicolumn{2}{c}{Flux (\jbm)\tablenotemark{a}} & \colhead{\%}\\
\cline{2-3}\\
\colhead{Source} & \colhead{B configuration} & \colhead{D configuration} & \colhead{resolved out}}
\startdata
\cutinhead{\mef}
IRc5 & 2.0 & 4.0 & 50\\
IRc6 & 1.3 & 1.4 & 0\\
\cutinhead{\dme}
IRc5 & 1.6 & 2.7 & 40\\
IRc6 & 1.4 & 1.4 & 0\\
\cutinhead{\etcn}
Hot Core & 2.8 & 3.3 & 0\\
IRc7 & 1.9 & 2.1 & 0\\
\cutinhead{\acetone}
HC-SW & 0.68 & 0.57 & 0\\
IRc7 & $<$0.5 & 0.25 & 0\\
\enddata
\tablenotetext{a}{Uncertainties are 20\% calibration uncertainty.}
\label{tab:resolve}
\end{deluxetable}
\clearpage

We present below an overview of the spectral line and imaging results for each molecule targeted in these observations.  The kinematics of the region will be discussed in Paper II (Friedel and Widicus Weaver, submitted).

\subsection{Individual Molecular Results}
\subsubsection{SiO}\label{sec:sio}
\begin{figure}[!ht]
\includegraphics[scale=0.9]{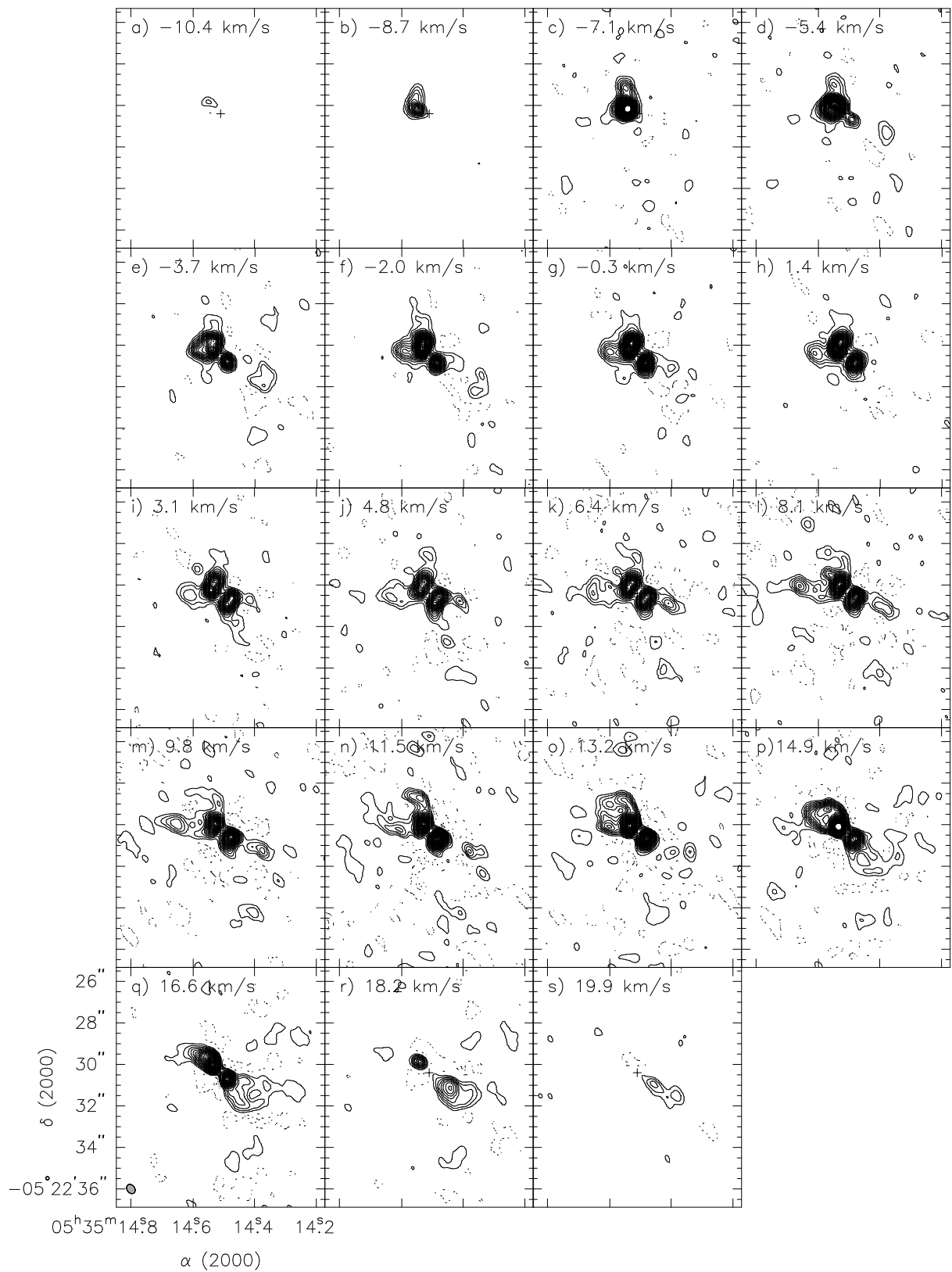}
\caption{\small{Velocity maps of the SiO $J=2-1$ transition toward Orion-KL. Each panel is labeled with the central rest velocity. Contours are $\pm6\sigma$, $\pm12\sigma$, $\pm18\sigma$,... ($\sigma$ = 42.8 m\jbm). The plus sign denotes the position of Source I, where the outflow originates. The beam size is shown in the lower left corner of panel (q).}\label{fig:sio-vel}}
\end{figure}

\begin{figure}[!ht]
\includegraphics[scale=0.9]{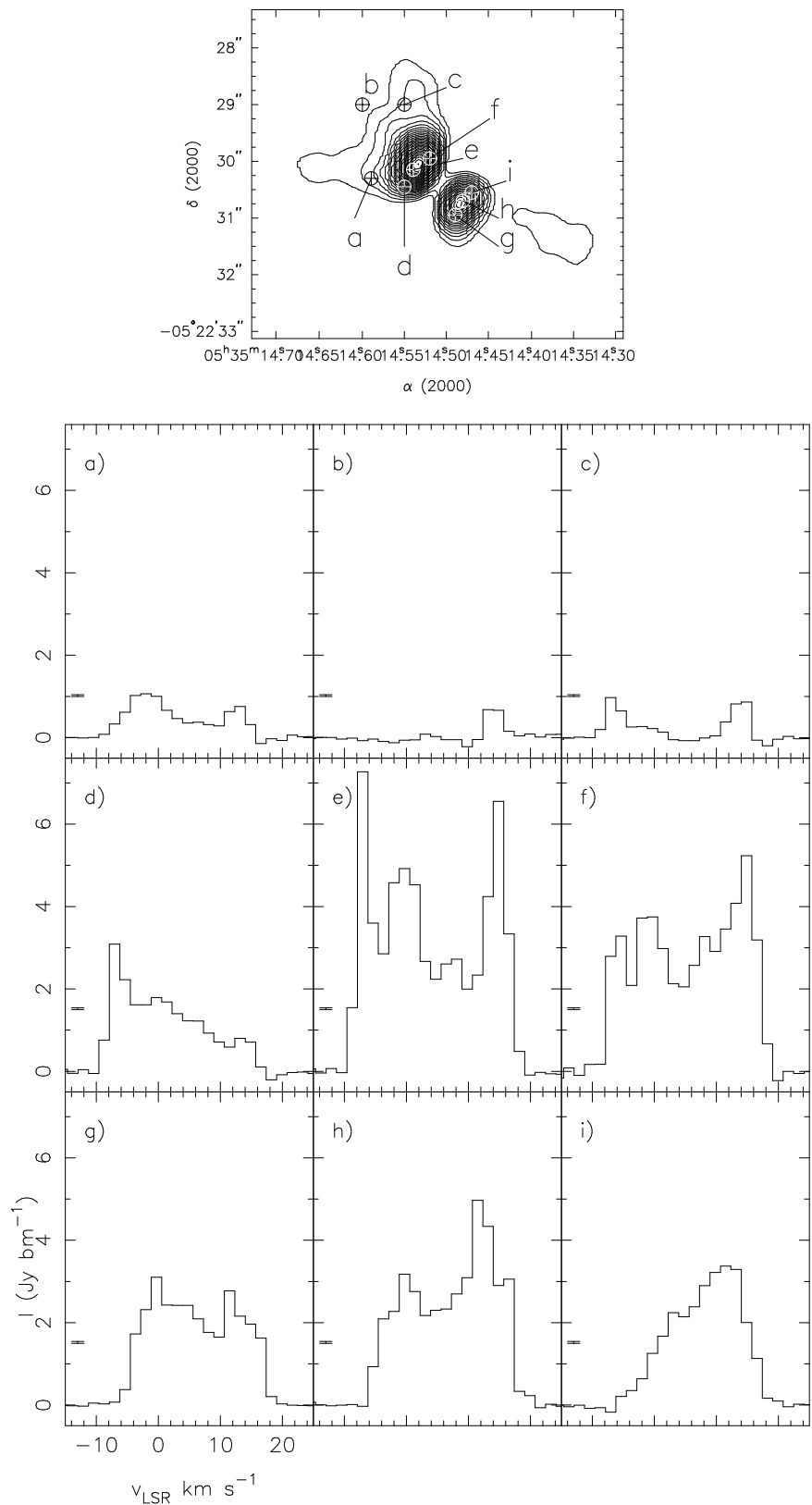}
\caption{\small{SiO emission in the Orion-KL region as observed in the A configuration. The top panel shows a map constructed from averaging all velocity channels. Contours are $\pm6\sigma$, $\pm9\sigma$, $\pm12\sigma$,... ($\sigma$ = 4.5 m\jbm). The marks and letters denote where spectra were taken, and the corresponding spectra are shown in panels (a) -- (i). The ``I'' bars denote the rms noise level in each spectrum.}\label{fig:sio}}
\end{figure}

We observed lines of SiO because it is an excellent shock tracer and thus a good indicator of some of the physical conditions of the region.  Emission from the $J=2-1$ transition of SiO is shown in Figure~\ref{fig:sio-vel}. Each panel shows a different velocity slice (from -10.4 to 19.9 \kms) with the velocity labeled in the upper left corner. Figure~\ref{fig:sio} shows a map of all of the SiO channels averaged together, along with spectra corresponding to different spatial locations in the source. The letters on the map denote which spectra match each physical location. From the double-lobed appearance of the emission, SiO seems to trace an outflow (see \citet{plambeck09} for a more detailed description of the outflow). This distribution is similar to the velocity integrated map presented by \cite{blake96} of a higher energy transition of SiO.  This outflow appears to be nearly in the plane of the sky since there is little velocity offset between the red- and blue-shifted lobes (less than a few \kms) when compared to the velocity dispersion within each lobe ($>$20 \kms). These results suggest that the SiO emission comes from a cone of emission rather than a solid outflow, indicated by the double-peaked spectra seen in both lobes (e.g. panels (e) and (h) from Figure~\ref{fig:sio}). This is also supported by observations of CH$_3$CN, which shows a cavity in the emission matching the location of the central part of the cone \citep{wang10}.

The outflow is asymmetric, with the spectral lines associated with the northeast lobe having velocities between -7.0 and 14.9 \kms, and lines associated the southwest lobe having velocities between -0.3 and 11.5 \kms. The asymmetry may come from the fact that the southwest lobe is interacting with a larger amount of ambient material that is hindering its expansion. This will be investigated in more detail in Paper II. There is also a third spectral peak visible in the northeast lobe at -0.3 \kms. The origin of this feature is not known at this time. The center of the outflow is Source I, and the position angle of the outflow is roughly perpendicular to the SiO maser disk reported by \citet{wright95}. \citet{wright95} also noted that, given the brightness temperature, the ground state SiO emission appears to be weakly masing. These observations agree with this conclusion, as the peak brightness temperatures are in excess of 6000 K.

\clearpage

\subsubsection{Methyl Formate [\mef] \& Dimethyl Ether [\dme]}\label{mef}
\begin{figure}[!ht]
\includegraphics[angle=270,scale=0.85]{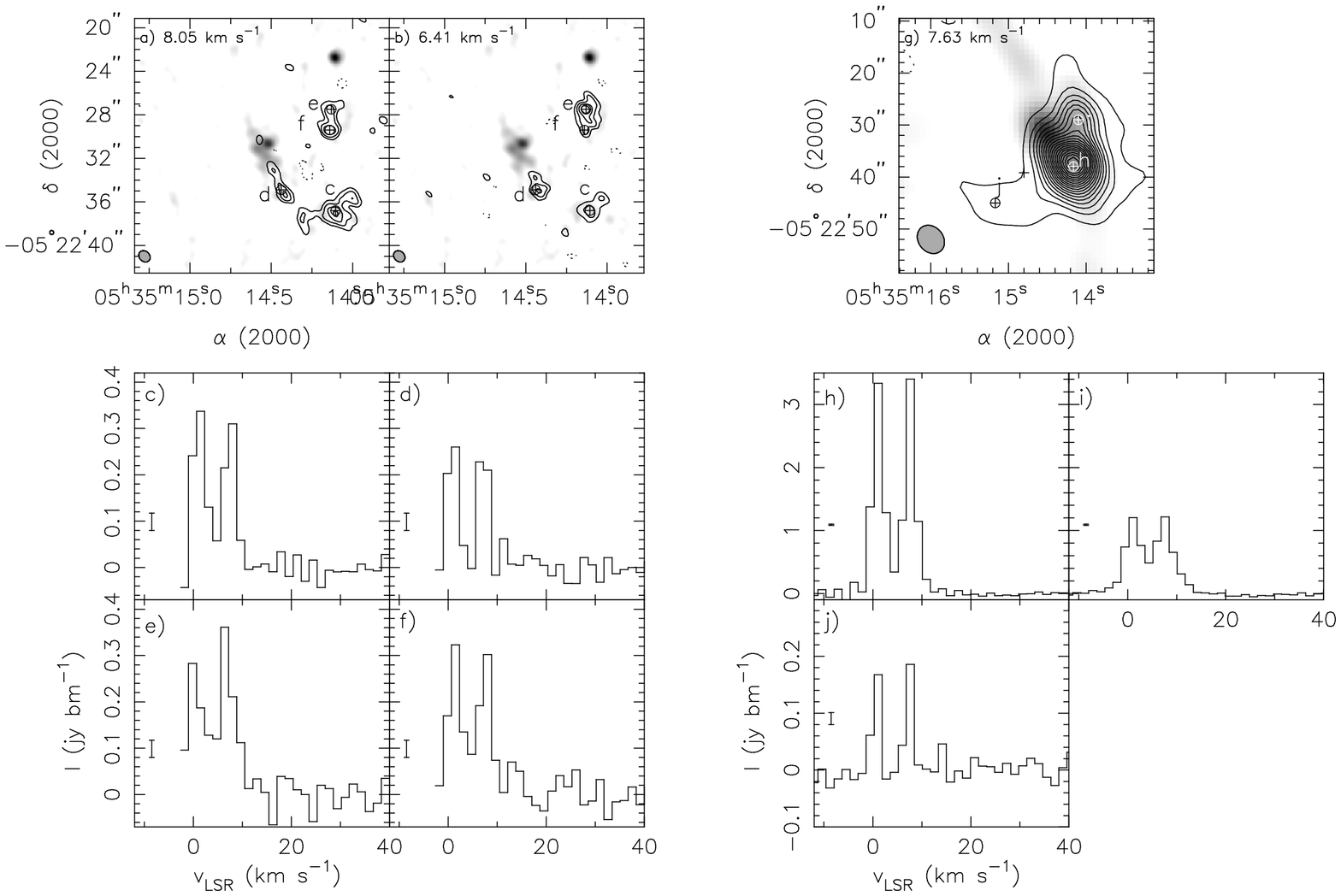}
\caption{\small{Methyl formate [\mef] emission from the Orion-KL region. Maps of the CARMA B configuration results at a) \vlsr = 8.05, and b) 6.41 \kms\ are shown with contours at $\pm$3$\sigma$, $\pm$5$\sigma$, $\pm$7$\sigma$, ...; $\sigma$ = 37.5 m\jbm.  The spectra corresponding to these maps are shown in panels (c) -- (f). The spectral panel letters correspond to the location labels on the map. The spectra show the two \mef\ peaks (A and E torsional states).  The ``I'' bar on the left hand side of each panel denotes the rms noise level.  Panel (g) shows the map of the CARMA D configuration results with contours at $\pm$5$\sigma$, $\pm15\sigma$, $\pm25\sigma$, ...; $\sigma$ = 20.7 m\jbm. The spectra from this map are shown in panels (h) -- (j), also with panel letters corresponding to the location labels on the map.} \label{fig:mef}}
\end{figure}

Observations of methyl formate [\mef] and dimethyl ether [\dme] reveal that these two molecules are co-spatial on all spatial scales.  This is in agreement with the lower spatial resolution results presented by \citet{neill11}. The discussion here will focus on \mef, but all conclusions also apply to \dme. Methyl formate emission has traditionally been associated with the Compact Ridge, where it was thought to be ablated off of grain surfaces by an outflow from the Hot Core \citep{blake87,liu02}. More recently, \citet{friedel08} suggested that the main \mef\ emission may arise from other more compact sources, rather than the Compact Ridge. Figure~\ref{fig:mef} shows the B and D configuration results, with \mef\ contours overlaid on the associated high-resolution continuum image (in gray scale).  No \mef\ emission was detected above the 3$\sigma$ (50.4 m\jbm) threshold in the A configuration observations, suggesting that this molecule is extended compared to the small A configuration beam.

The peaks of the \mef\ flux observed in both the high- and low-resolution images are located in approximately the same position. One exception is the peak observed in the D configuration image, located to the southeast of the main emission peak. This flux is completely resolved out by the higher-resolution observations, indicating that it is extended ($\geq$4.2\arcsec\, $>$1700 AU). Table~\ref{tab:resolve} lists the peak flux measured for each molecule at the positions of the main compact sources.  In the case of IRc5, it appears that 50\% of the flux is being resolved out by the high-resolution observations, indicating that this emission is coming from a region at least 4.5\arcsec\ in size. However, for IRc6 it appears that no flux is being resolved out. Based on the non-detection of \mef\ in the A configuration observations presented here, it appears that the main \mef\ emission arises from regions that are at least 2.4\arcsec\ (1000 AU) in size.

\begin{figure}[!ht]
\includegraphics[angle=270,scale=0.9]{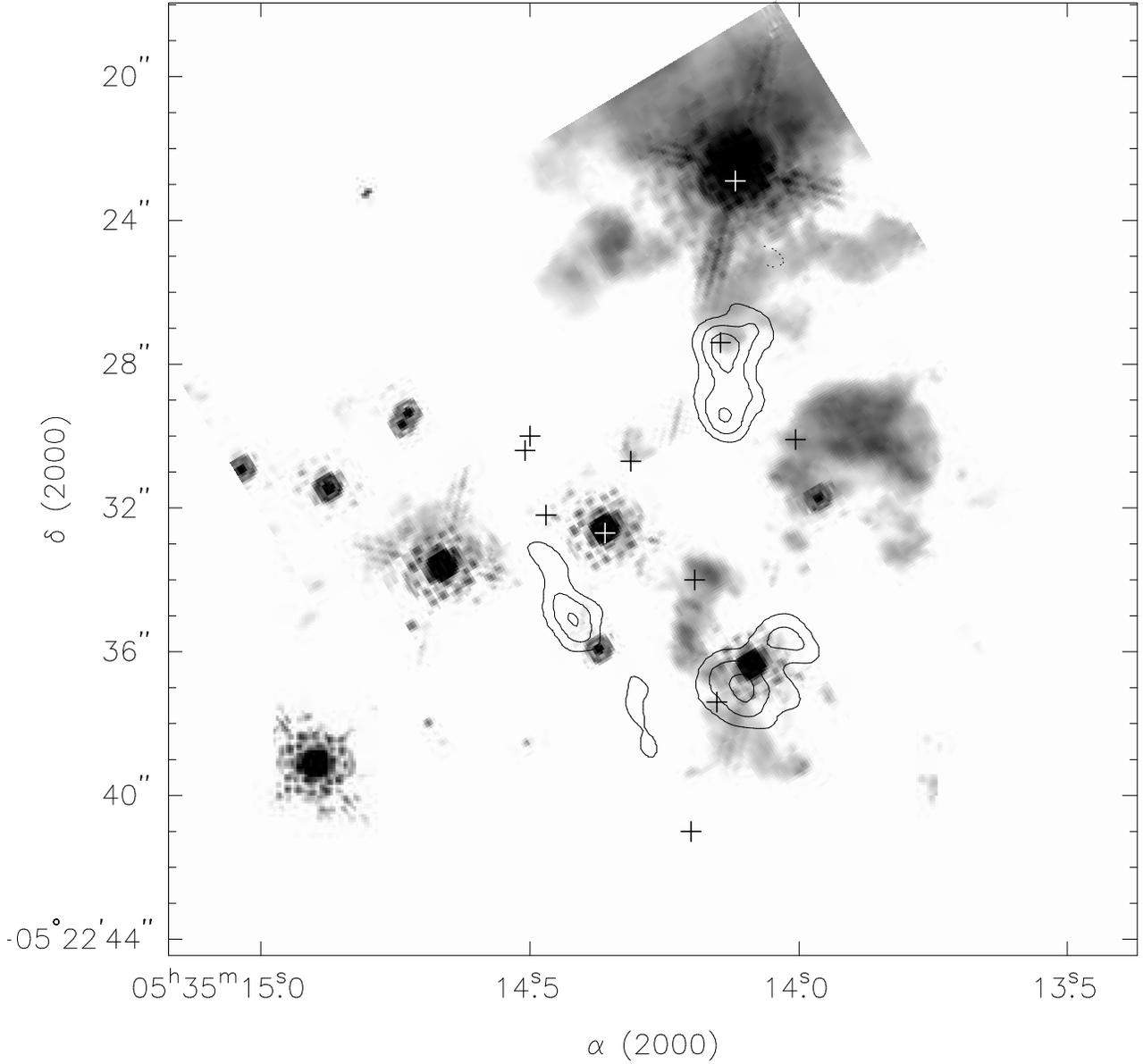}
\caption{\small{Methyl formate [\mef] emission from the B configuration overlaid on 2 $\mu$m gray scale \citep{friedel11}.. The map is an average moment map of both peak velocity channels and the contours are $\pm$3$\sigma$, $\pm$5$\sigma$, $\pm$7$\sigma$, ...; $\sigma$ = 37.5 m\jbm. The plus signs denote the following sources (from E to W: Source I, hot core, SMA1, Source n, IRc7, compact ridge, IRc4, IRc5, IRc6, BN, and IRc3.} \label{fig:h-mef}}
\end{figure}

Figure~\ref{fig:h-mef} shows the combined moment map of the B configuration \mef\ data overlaid on a 2 $\mu$m continuum, highlighting three main regions of compact emission. The first region is between IRc6 and IRc20, with emission peaks having different velocities near each source (see Figure~\ref{fig:mef}). The second region of \mef\ emission is a double-lobed feature centered on star k. The third region of \mef\ emission is located very near star p and extends toward SMA1.  None of these regions are associated with the nominal position of the Compact Ridge.

\begin{figure}
\includegraphics[angle=90,scale=1.1]{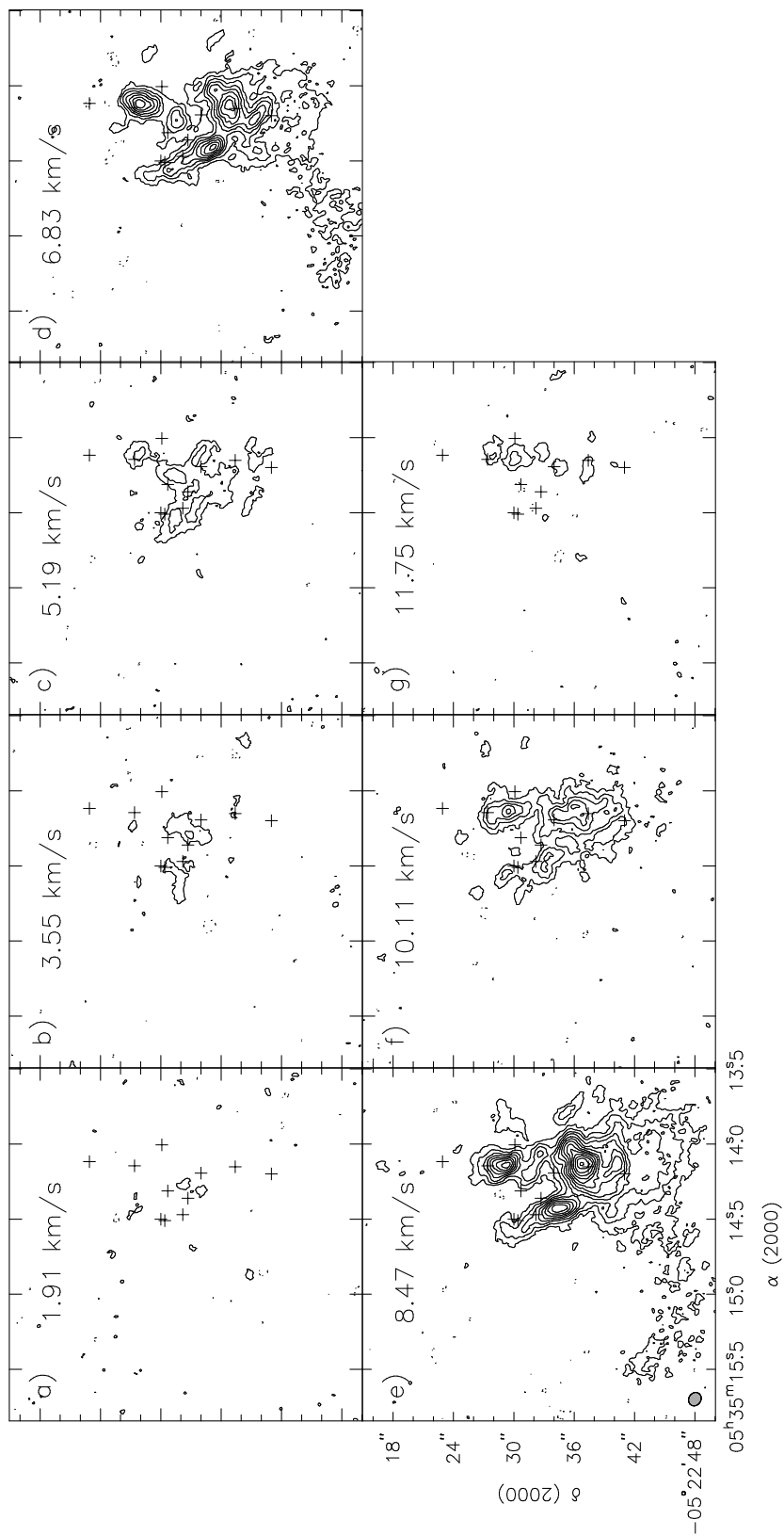}
\caption{\small{Naturally weighted maps of \mef\ using all available data. The contours are $\pm3\sigma$, $\pm6\sigma$, $\pm9\sigma$, ... ($\sigma$ = 12.6 m\jbm). The synthesized beam is in the lower left corner of panel (e). The channel velocity of each panel is given in the upper center of each panel. Plus signs are the same as in Figure~\ref{fig:h-mef}.} \label{fig:mef-vel}}
\end{figure}

Figure~\ref{fig:mef-vel} shows all B and D configuration \mef\ data inverted together with natural weighting. Each panel is labeled with the corresponding \vlsr\, and the synthesized beam is in the lower left corner of panel (e). These maps illustrate the full distribution of \mef. Methyl formate is detectable from 1.9 \kms\ to 11.75 \kms, with the three aforementioned peaks, and the extended structure toward the southeast. \cite{blake96} and \cite{beuther05} observed higher energy transitions of \mef\ with slightly lower spatial resolution. Their data also indicate a weak \mef\ extension toward the hot core, however due to more sparse {\it u-v} sampling, higher noise, and velocity averaging, many of the details of the emission features are not clearly detected. It is interesting to note that the bulk of the emission between the three transitions are very similar, even though there is a factor of $\sim$5-15 difference in $E_u$ and $\sim$2.5-3 difference in $S\mu^2$.

\clearpage

\subsubsection{Ethyl Cyanide [\etcn], Methyl Cyanide [\mtcn], \& Vinyl Cyanide [\vycn]}\label{etcn}
\begin{figure}[!ht]
\includegraphics[angle=270,scale=1.2]{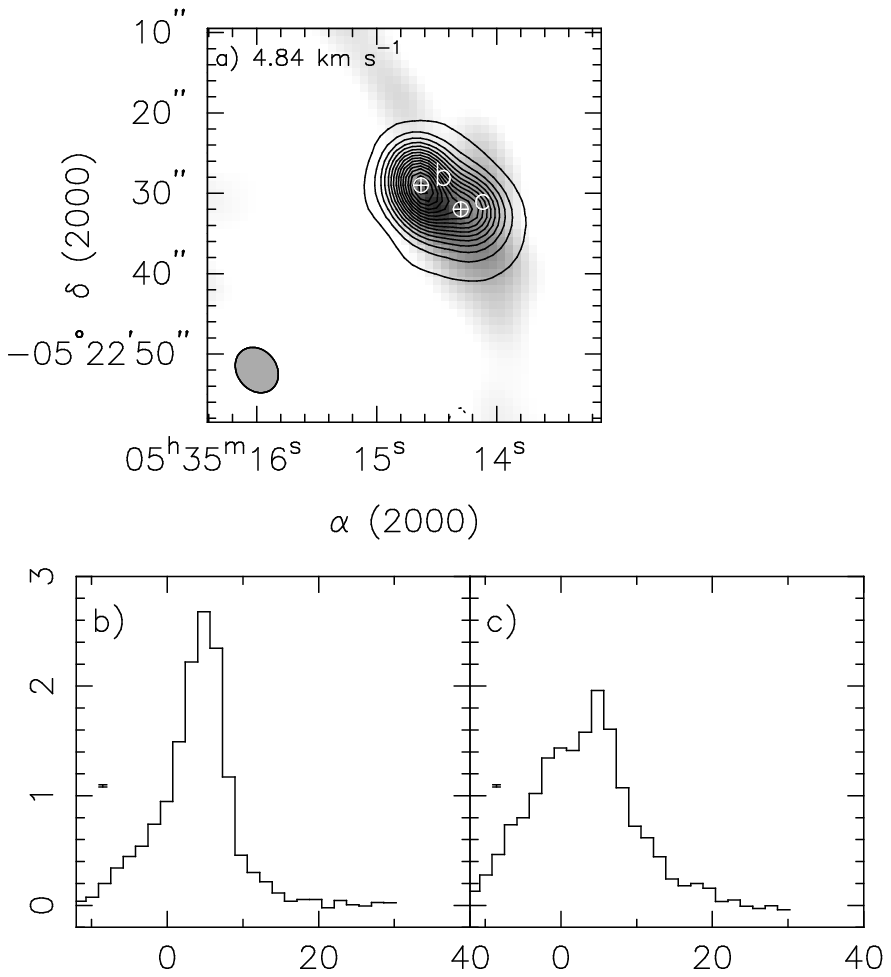}
\caption{\small{Ethyl cyanide [\etcn] emission from the Orion-KL region as observed in the D configuration. Panel (a) shows a map of the \vlsr\ = 4.85 \kms\  emission with contours of $\pm$5$\sigma$, $\pm15\sigma$, $\pm25\sigma$, ... ($\sigma$ = 20.7 m\jbm). The gray scale is the $\lambda$ = 3 mm continuum. The spectra from this map are shown in panels (b) and (c), with the letters on the map denoting the location of the spectra. The ``I'' bar denotes the rms noise level on each spectrum.} \label{fig:etcnD}}
\end{figure}

\begin{figure}[!ht]
\includegraphics[angle=270,scale=1.2]{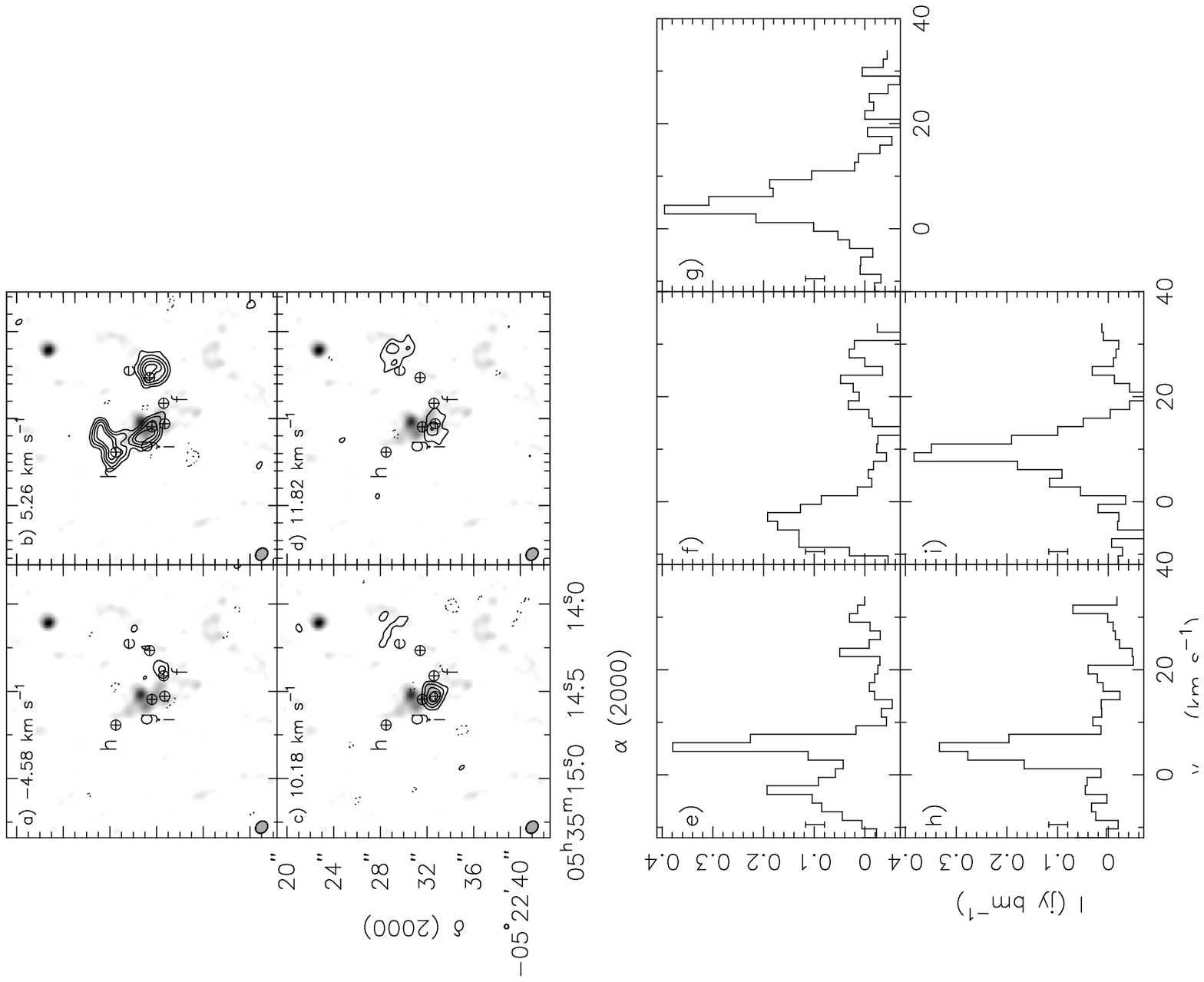}
\caption{\small{Ethyl cyanide [\etcn] emission from the Orion-KL region as observed in the B configuration. Panels (a) -- (d) show maps of the \vlsr\ = -4.58, 5.26, 10.18, and 11.82 \kms\ emission, respectively, with contours of $\pm$3$\sigma$, $\pm$5$\sigma$, $\pm$7$\sigma$, ... ($\sigma$ = 37.3 m\jbm). The gray scale is the $\lambda$ = 3 mm continuum. The spectra from these maps are shown in panels (e) -- (i). The panel letters denote the location of the spectra on the map, and the ``I'' bar denotes the rms noise level.} \label{fig:etcnB}}
\end{figure}

\begin{figure}[!ht]
\includegraphics[angle=270,scale=1.2]{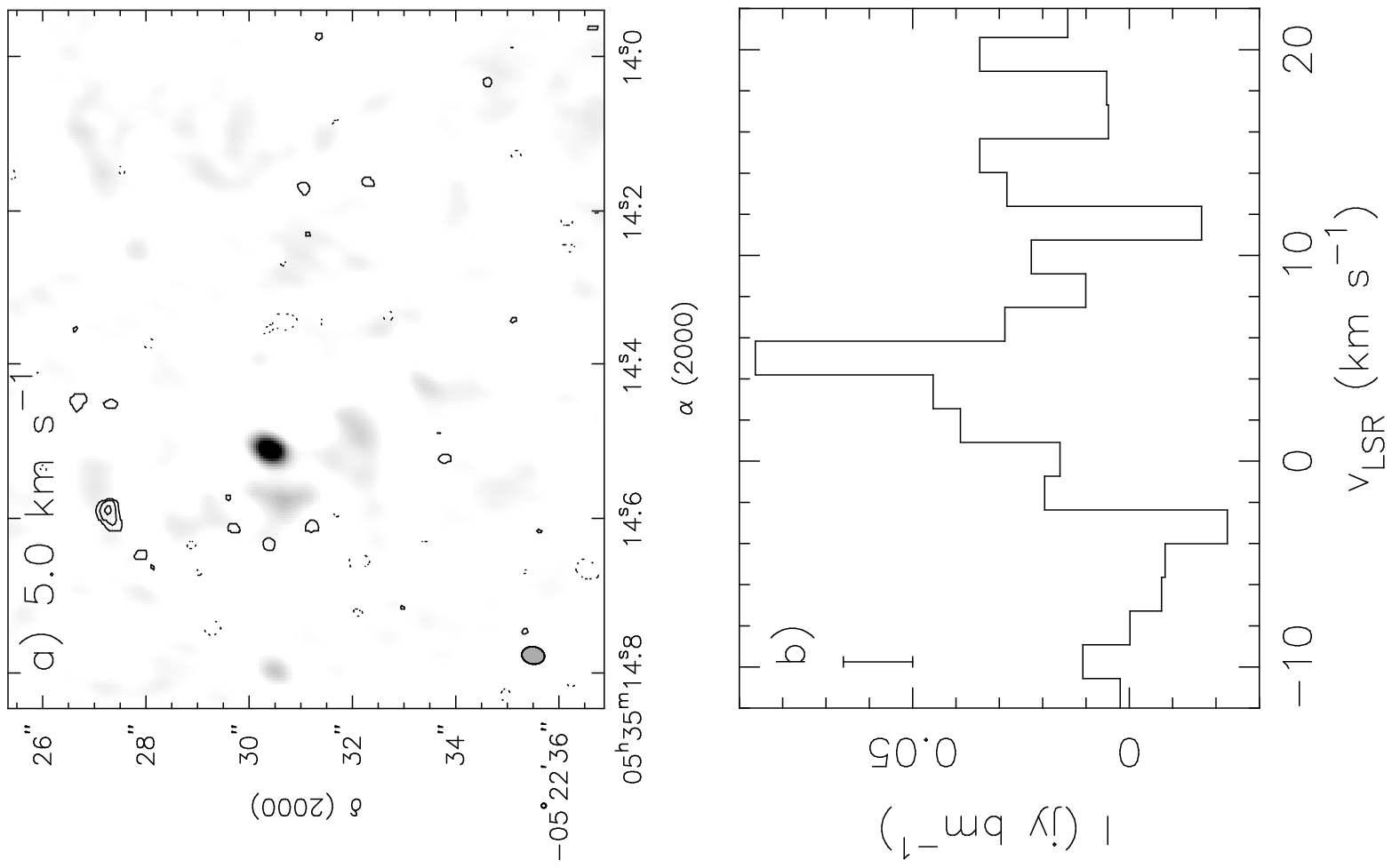}
\caption{\small{Ethyl cyanide [\etcn] emission from the Orion-KL region as observed in the A configuration. Panel (a) shows a map of the  \vlsr\ = 5.00 \kms emission with contours of $\pm$3$\sigma$, $\pm$4$\sigma$, $\pm$5$\sigma$, ... ($\sigma$ = 16.8 m\jbm). The gray scale is the $\lambda$ = 3 mm continuum. Panel (b) shows the spectrum from the strongest emission peak.} \label{fig:etcnA}}
\end{figure}

Observations of ethyl cyanide [\etcn], methyl cyanide [\mtcn], and vinyl cyanide [\vycn] reveal that these molecules are co-spatial at all observed spatial resolutions. The discussion here will focus on \etcn, but all conclusions also apply to \mtcn\ and \vycn.  Traditionally, \etcn\ has been associated with the Hot Core region \citep{blake87}. \citet{friedel08} suggested that the \etcn\ emission comes from multiple sources in and near the Hot Core and IRc7. Figures~\ref{fig:etcnD}, \ref{fig:etcnB}, and \ref{fig:etcnA} show the CARMA D, B, and A configuration results, respectively. The labels on each map denote which spectrum is associated with which location.  The spectra from the B and D configurations observations show that there are multiple velocity components at several of the emission peaks. However, these multiple components are not clearly seen in the A configuration results. Comparison of the flux between the B and D configuration observations (see Table~\ref{tab:resolve}) shows that, within the 20\% calibration uncertainties, no flux is resolved out by the smaller B configuration synthesized beam. However, only a small fraction of the total flux is detected by the subarcsecond A configuration beam. This indicates that while the \etcn\ emission is compact (200-300 AU), most of the gas is on the scale of a few arcseconds (1000-1700 AU).

\begin{figure}[!ht]
\includegraphics[angle=0,scale=0.8]{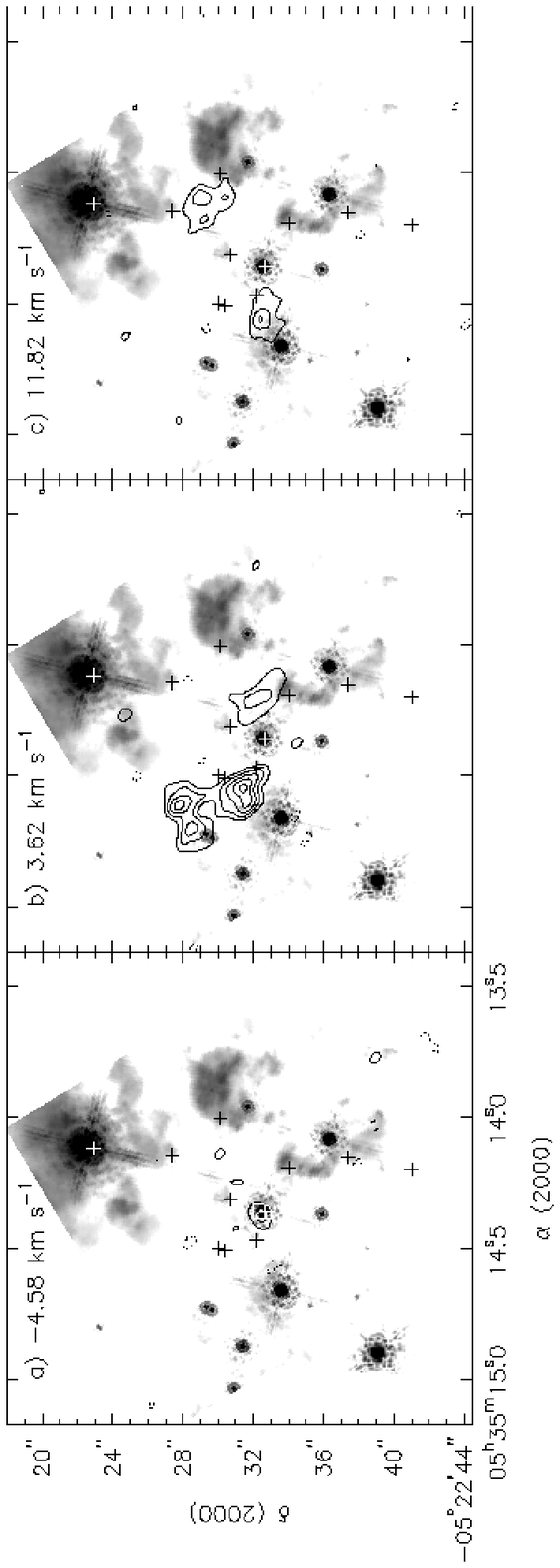}
\caption{\small{Ethyl cyanide [\etcn] emission from the B configuration overlaid on 2 $\mu$m gray scale. The contours are $\pm$3$\sigma$, $\pm$5$\sigma$, $\pm$7$\sigma$, ...; $\sigma$ = 37.3 m\jbm. Each panel shows a different velocity channel: a) -4.58 \kms, b) 3.62 \kms, and c) 11.82 \kms . Plus signs are the same as in Figure~\ref{fig:h-mef}.} \label{fig:h-etcn}}
\end{figure}

Figure~\ref{fig:h-etcn} shows the B configuration \etcn\ results from several velocities overlaid on the 2 $\mu$m grayscale continuum. In Figure~\ref{fig:h-etcn}a, the contours are centered very close to star n. NH$_3$ emission has also been detected toward this source at a similar rest velocity ($\sim$-4 \kms) by \citet{min89}. This indicates that the region has a high density, and it is possible that there is a dust shell or torus surrounding the star \citep{simp06,min89,green04}. Figure~\ref{fig:h-etcn}b shows several strong concentrations of \etcn\ emission, the strongest of which is the peak to the east of the Hot Core and SMA1. Also of note are the peaks near IRc7/4 and CB4, which are the only areas where \etcn\ is detected in the A configuration.  Figure~\ref{fig:h-etcn}c shows the two peaks reported by \cite{friedel08} at a similar \vlsr\ ($\sim$11 \kms). These unusual peaks have now been observed in multiple data sets and for multiple transitions of \etcn, so it is quite likely that they are real emission features and not a side-product of the data reduction process. It should be noted that the feature observed near IRc3/6/20 is coincident with peaks of \mef\ and \dme\ and occurs at a similar rest velocity.

\begin{figure}
\includegraphics[scale=0.9]{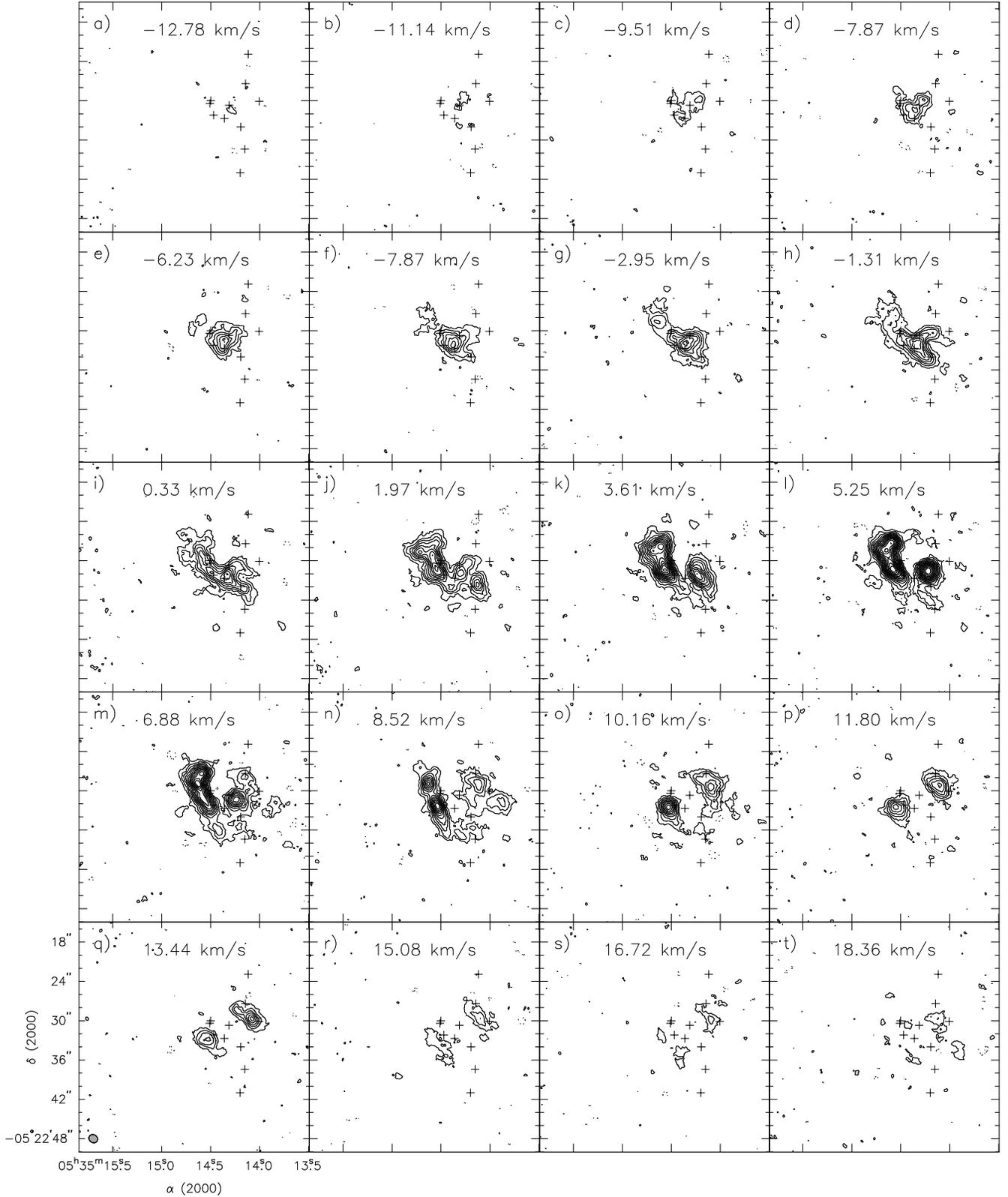}
\caption{\small{Naturally weighted maps of \etcn\ using all available data. The contours are $\pm3\sigma$, $\pm6\sigma$, $\pm9\sigma$, ... ($\sigma$ = 12.6 m\jbm). The synthesized beam is in the lower left corner of panel (q). The channel velocity of each panel is given in the upper center of each panel. Plus signs are the same as in Figure~\ref{fig:h-mef}.} \label{fig:etcn-vel}}
\end{figure}

Figure~\ref{fig:etcn-vel} shows all A, B, and D configuration \etcn\ data inverted together with natural weighting. Each panel is labeled with the corresponding \vlsr\, and the synthesized beam is in the lower left corner of panel (q). These maps show that the full distribution of \etcn. \etcn\ is detectable from -12.8 \kms\ to 18.4 \kms, which is a much larger velocity spread than suggested by the traditional Hot Core models \citep[e.g.][]{blake87}. Of particular note are the negative velocity maps, as these velocities have traditionally been associated with the Extended Ridge (a much larger structure that is resolved out by these observations), and not compact structures \citep[e.g][]{blake87}. Additionally, the peaks at 10-14 \kms\ show that there are two additional \etcn\ concentrations, one near the Hot Core-SW and the other near IRc6. Similarly to the \mef\ observations, \cite{blake96} and \cite{beuther05} also observed higher energy \etcn\ transitions. The maps presented in their papers have similar intensity peak positions, but much of the finer detail shown here is not seen. This may be due to the more complete {\it u-v} sampling and lower noise level of the current observations, and/or an actual difference in the distribution of the different transitions, as the higher energy transitions preivously are $\sim$4-12 times higher in $E_u$ and $\sim$3-4.8 times higher in $S\mu^2$.

\clearpage

\subsubsection{Acetone [\acetone]}\label{acetone}
\begin{figure}[!ht]
\includegraphics[angle=0,scale=0.9]{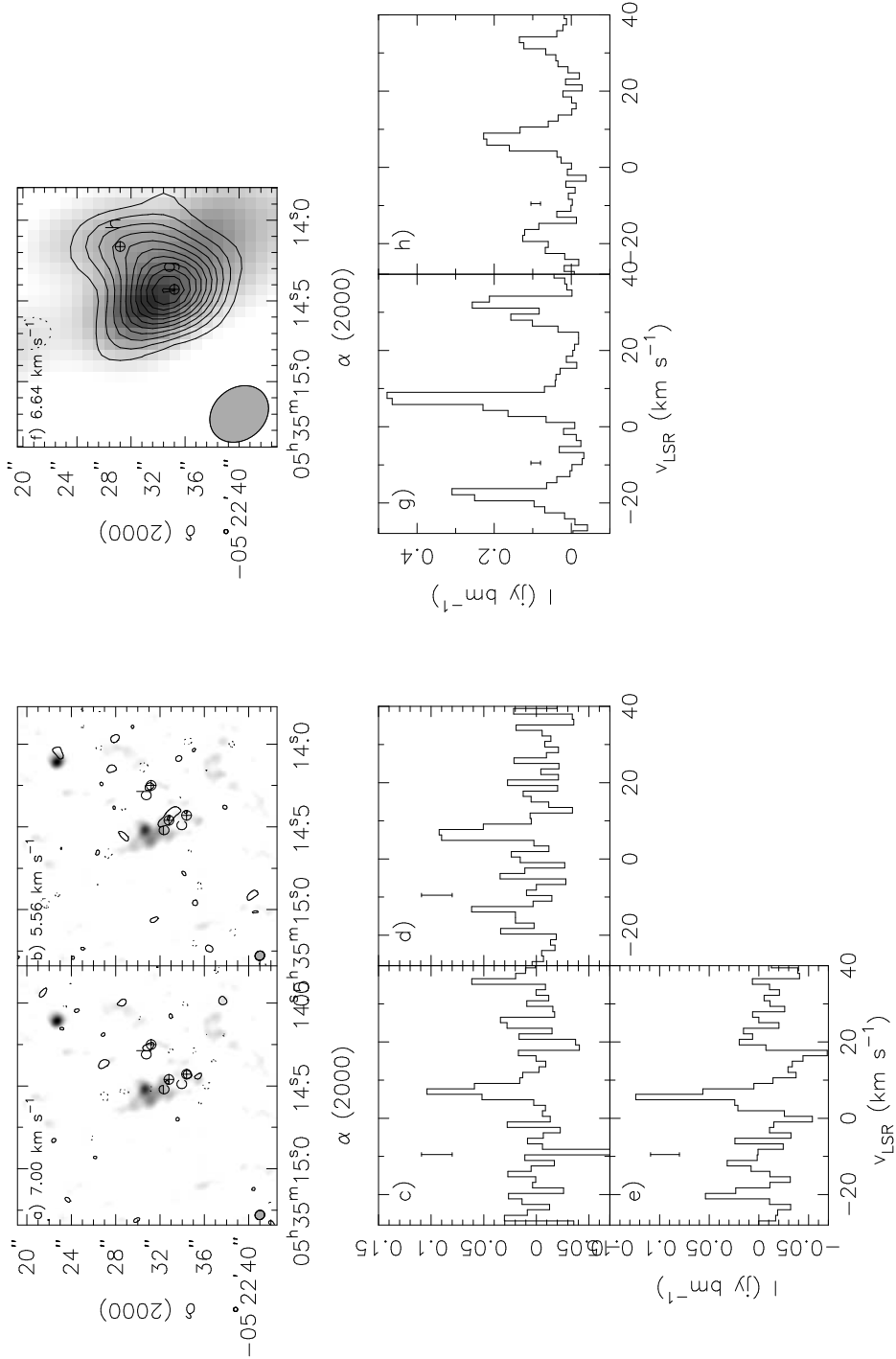}
\caption{\small{Acetone [\acetone] emission from the Orion-KL region. Panels (a) and (b) show maps of the CARMA B configuration results at \vlsr = 7.00 and 5.56 \kms\, respectively, with contours at $\pm$3$\sigma$, $\pm$5$\sigma$, $\pm$7$\sigma$, ...; $\sigma$ = 29.1 m\jbm. The spectra shown in panels (c) -- (e) correspond to the letters on the maps.  Panel (f) shows the map of the CARMA D configuration results with contours at $\pm$3$\sigma$, $\pm5\sigma$, $\pm7\sigma$, ...; $\sigma$ = 24.7 m\jbm. The corresponding spectra are shown in panels (g) and (h). The gray scale is the continuum from the respective array configuration. Note that the \acetone\ spectrum is a triplet of lines arising from the torsional splitting.} \label{fig:ace}}
\end{figure}

\citet{friedel08} found that \acetone\ was only present in regions where both large N-bearing molecules (e.g. \etcn) and large O-bearing molecules (e.g. \dme) appeared to be co-spatial. Figure~\ref{fig:ace} shows the \acetone\ emission from the Orion-KL region as observed in the B and D configurations. Note that in addition to the \acetone\ emission peak in the region where \etcn\ and \dme\ are co-spatial, there is another region of emission near IRc6 that is seen in the D configuration map (spectrum h).

\begin{figure}[!ht]
\includegraphics[angle=270,scale=0.8]{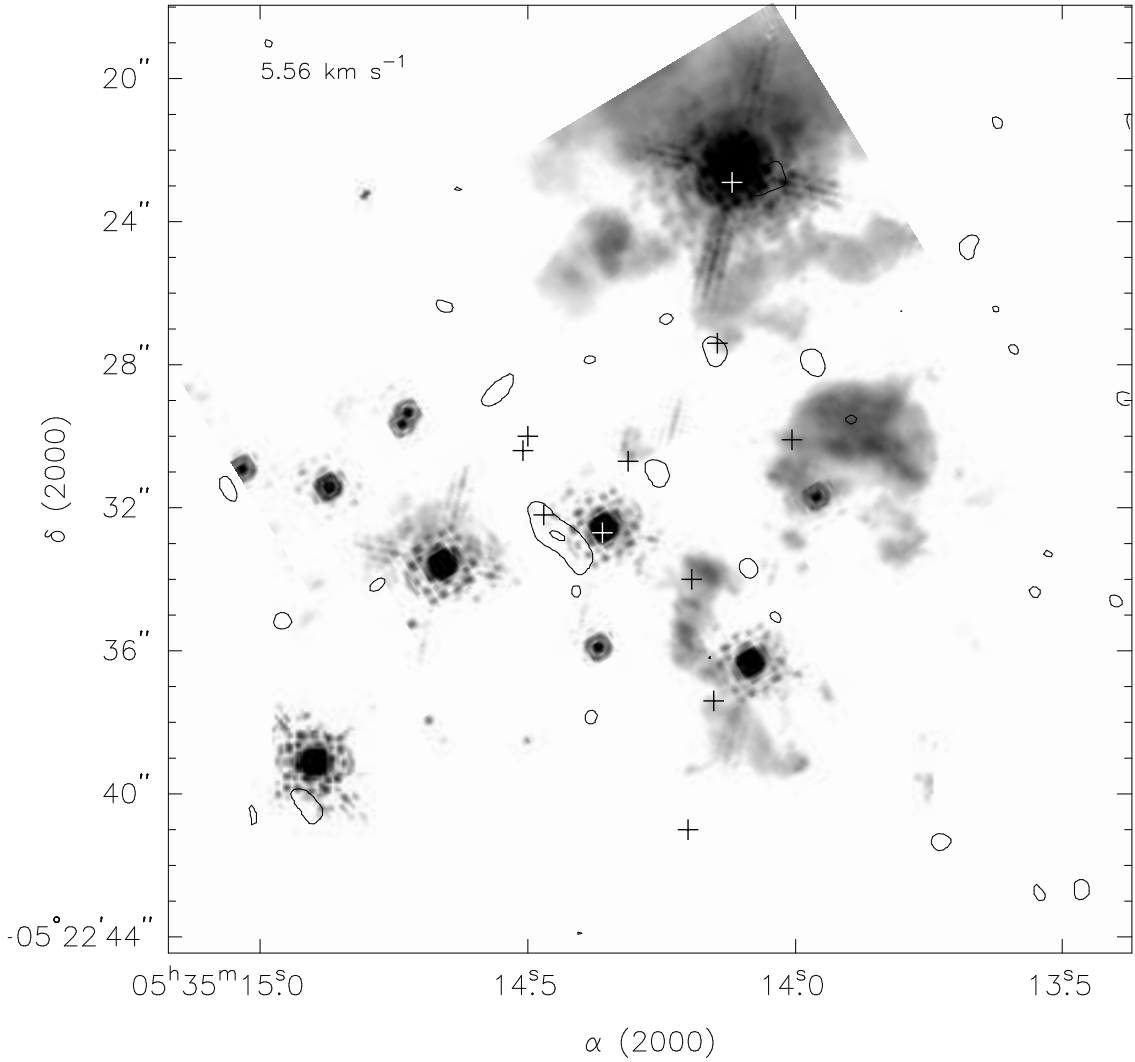}
\caption{\small{Acetone [\acetone] emission at \vlsr\ = 5.56 \kms\ as observed in the B configuration observations, overlaid on 2 $\mu$m gray scale. The contours are $\pm$3$\sigma$, $\pm$5$\sigma$, $\pm$7$\sigma$, ...; $\sigma$ = 29.1 m\jbm. Plus signs are the same as in Figure~\ref{fig:h-mef}.} \label{fig:h-ace}}
\end{figure}

Comparing the flux between the high- and low-resolution observations (see Table~\ref{tab:resolve}) shows that, within the 20\% calibration uncertainties, no flux is resolved out by the smaller B configuration synthesized beam\footnote{While different transitions of \acetone\ were observed by each configuration, the upper state energies and line strengths of both transitions are close enough to make the comparison meaningful.}. This indicates that all of the \acetone\ flux from IRc7 and the Hot Core regions is compact. The emission from the IRc6 region is only detected by the larger D configuration synthesized beam. Therefore, the emission in this location is either too extended or too weak to be detected with the smaller beam. Since other molecules [\dme, \mef, and \etcn] have been detected toward this region at high resolution, it is more likely that the \acetone\ emission is too weak to be detected above the 3$\sigma$ noise cutoff (87.3 m\jbm). Figure~\ref{fig:h-ace} shows \acetone\ emission at \vlsr\ = 5.56 \kms\ from the B configuration observations overlaid on the 2 $\mu$m continuum gray scale.  The two previously identified regions of acetone are weakly detected near SMA1 and IRc7.

\begin{figure}
\includegraphics[scale=1.1]{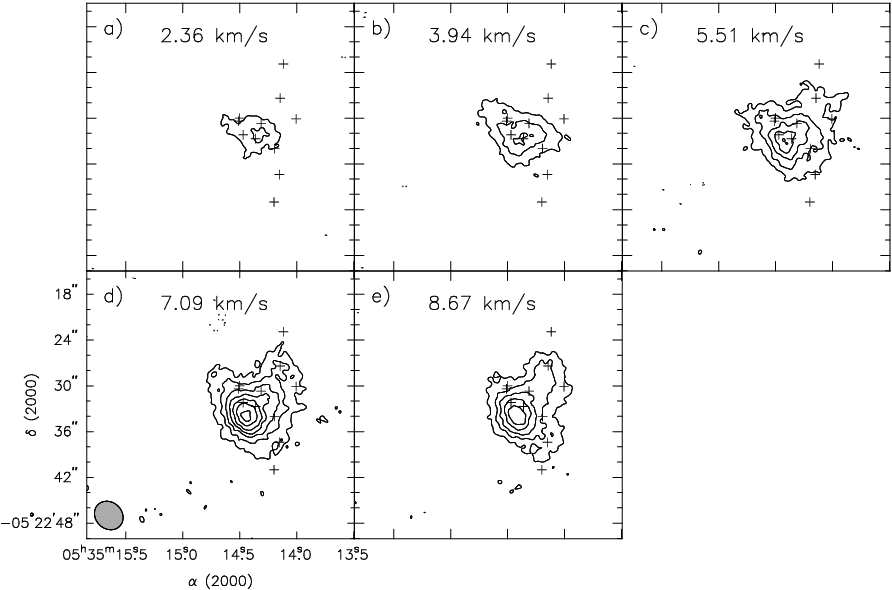}
\caption{\small{Naturally weighted maps of \acetone\ using all available data. The contours are $\pm3\sigma$, $\pm6\sigma$, $\pm9\sigma$, ... ($\sigma$ = 20.6 m\jbm). The synthesized beam is in the lower left corner of panel d. The channel velocity of each panel is given in the upper center of each panel. Plus signs are the same as in Figure~\ref{fig:h-mef}.} \label{fig:ace-vel}}
\end{figure}

Figure~\ref{fig:ace-vel} shows all B and D configuration data inverted together with natural weighting. Each panel has the \vlsr\ labeled in the center top and the synthesized beam is in the lower left corner of panel (d). These maps show that the full distribution of \acetone. \acetone\ is detectable from 2.4 \kms\ to 8.7 \kms, the narrowest velocity range of the species studied in this work. The synthesized beam for these maps is notably larger than that of the \etcn\ and \mef\ maps, thus much of the fine structure is not seen.

\clearpage

\subsubsection{Formic Acid [\fa] and Acetaldehyde [\acal]}\label{fa_acal}
\begin{figure}[!ht]
\includegraphics[angle=0,scale=1.1]{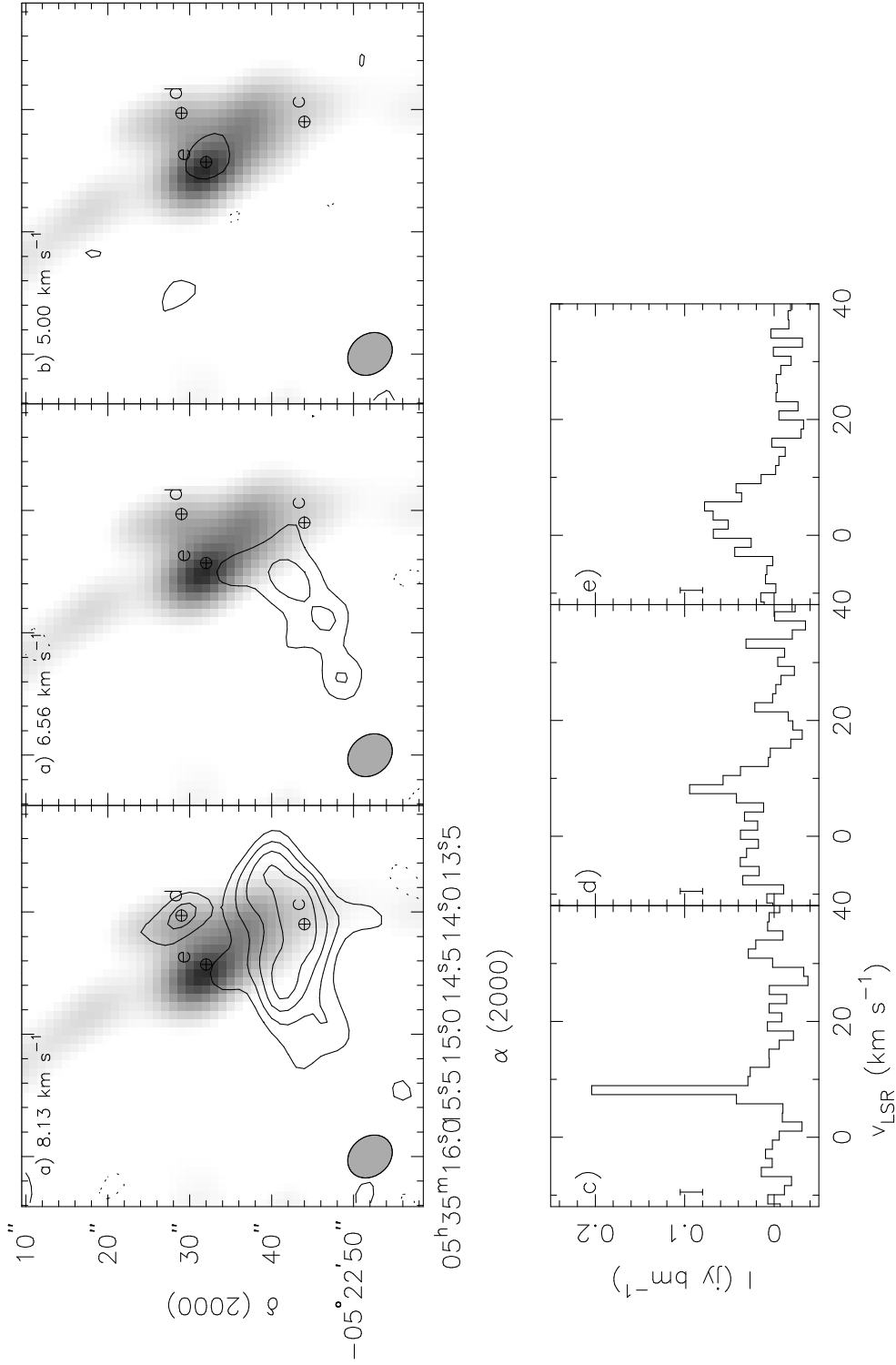}
\caption{\small{Formic acid [\fa] emission from the D configuration observations, at \vlsr\ = 8.13 and 5.00 \kms. Contours are $\pm$3$\sigma$, $\pm$5$\sigma$, $\pm$7$\sigma$, ...; $\sigma$ = 24.9 m\jbm. The spectra from these maps are shown in panels (c) -- (e), with the letters on the map denoting the location of the spectra. The gray scale is the 3 mm continuum.} \label{fig:fa}}
\end{figure}

\begin{figure}[!ht]
\includegraphics[angle=270,scale=0.8]{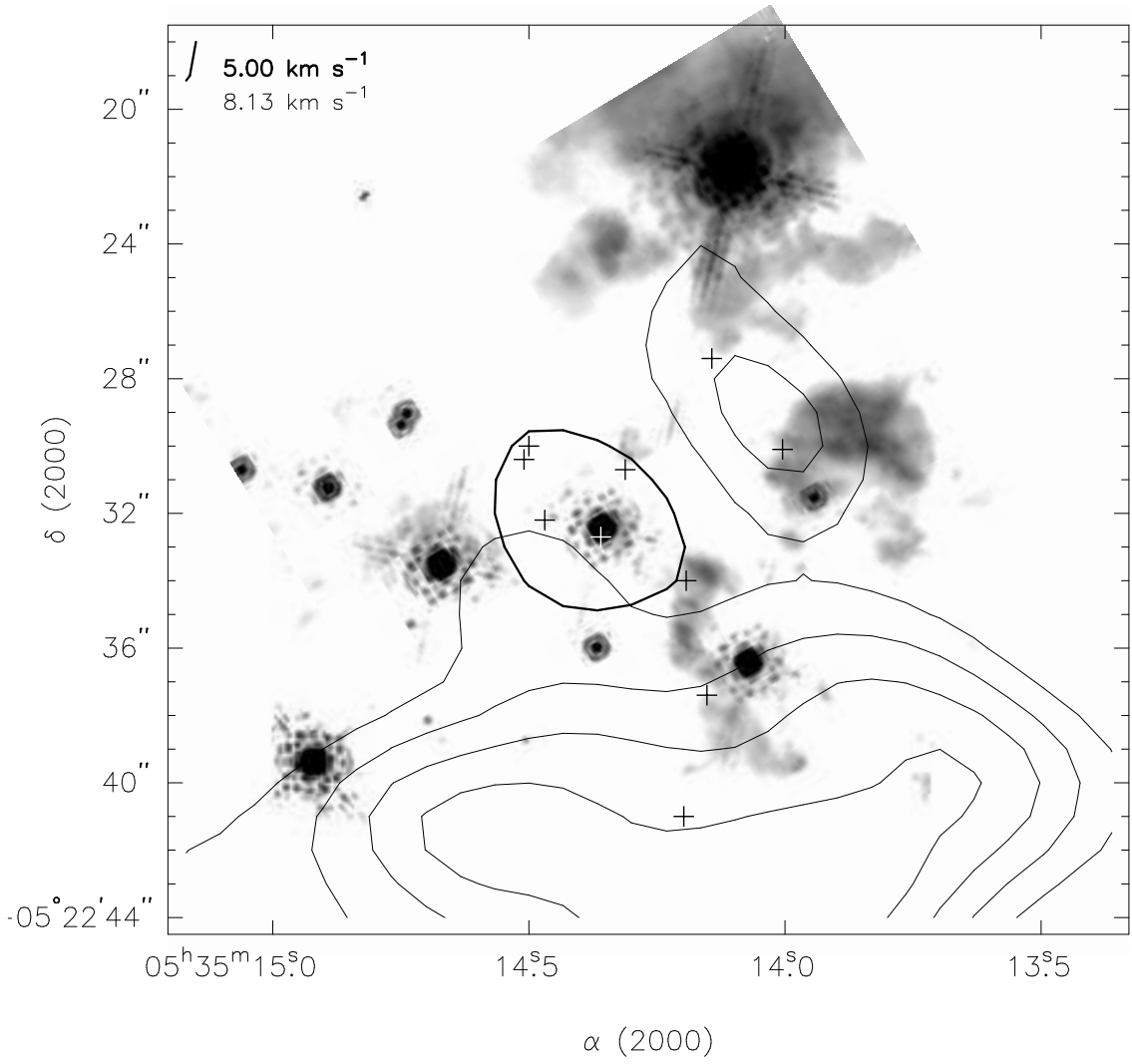}
\caption{\small{Formic acid [\fa] emission from the D configuration observations overlaid on the 2 $\mu$m gray scale. The contours are $\pm$3$\sigma$, $\pm$5$\sigma$, $\pm$7$\sigma$, ...; $\sigma$ = 24.9 m\jbm. Plus signs are the same as in Figure~\ref{fig:h-mef}.} \label{fig:h-fa}}
\end{figure}

\fa, originally detected in this region by \citet{liu02}, was included in the B configuration observations, but was not detected above the 3-$\sigma$ cutoff (144 m\jbm). It was detected in the D configuration observations,\footnote{While different \fa\ transitions were included in each array configuration, the upper state and energies and line strengths of the transitions are identical, thus they can be considered equivalent for the purpose of comparison.} and the resulting detection is shown in Figure~\ref{fig:fa}. There are two main peaks of emission for \fa:  one near the Compact Ridge that had been previously identified, and one near IRc6 at the traditional \vlsr\ for the Compact Ridge ($\sim$8 \kms) that has not been previously observed. There is also another previously unknown peak, with a wide line profile, at a \vlsr\ of $\sim$5 \kms\ near star n. This velocity has traditionally been associated with N-bearing species toward the Hot Core. The non-detection in the B configuration observations indicates that there are no compact concentrations of \fa\ in the region, but rather that the vast majority of its emission is extended (in regions of at least 4.2\arcsec\, $>$1700 AU). Figure~\ref{fig:h-fa} shows \fa\ emission from the D configuration overlaid on the 2 $\mu$m gray scale.

\begin{figure}[!ht]
\includegraphics[angle=0]{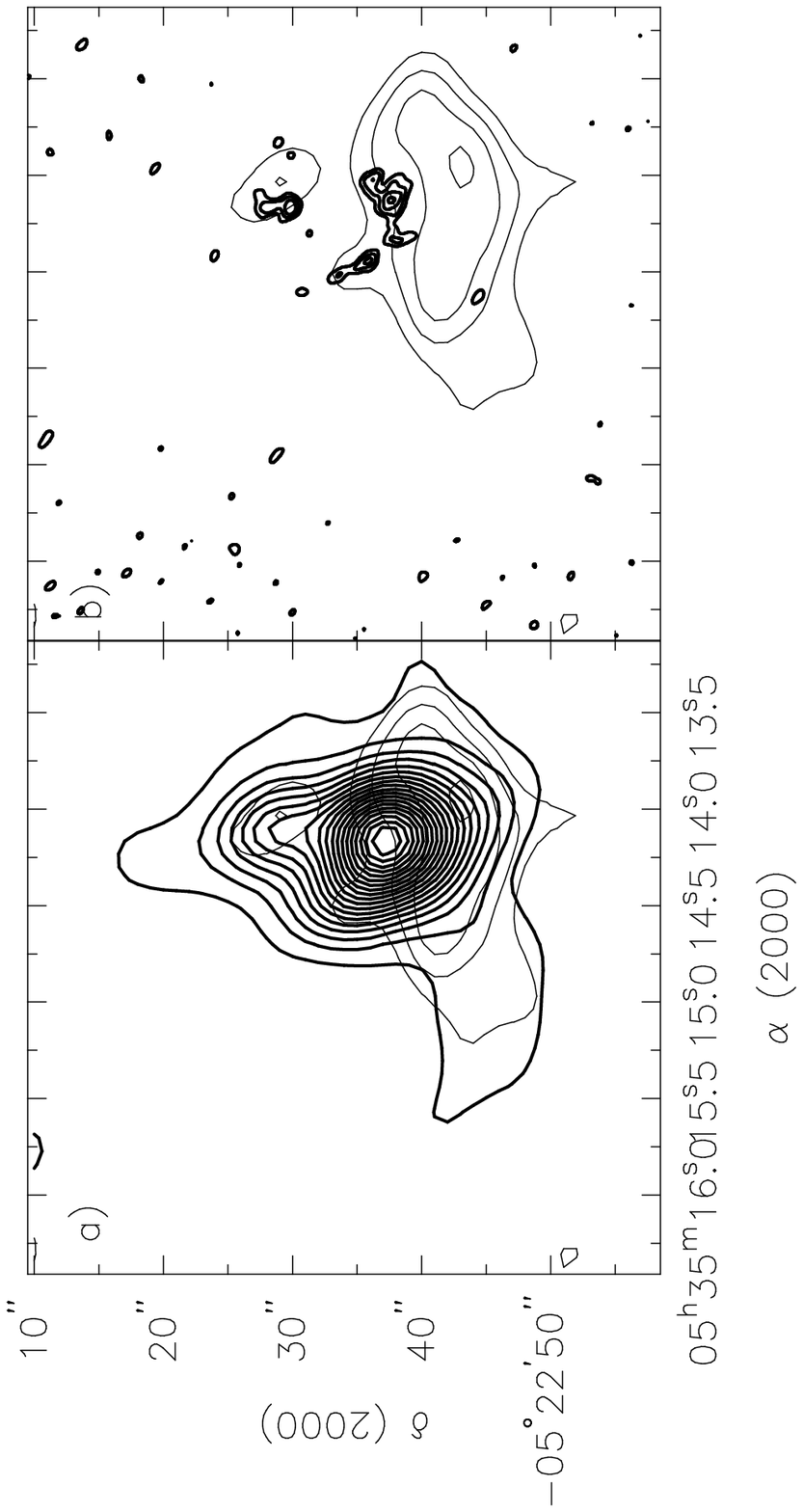}
\caption{\small{Comparison of \fa\ and \mef\ emission. The \fa\ emission is in thin contours and the \mef\ emission in heavy contours. Panel (a) shows the D configuration observations of \mef\ and \fa. Contours are $\pm$5$\sigma$, $\pm15\sigma$, $\pm25\sigma$, ...; $\sigma$ = 20.7 m\jbm\ for \mef\ and $\pm$3$\sigma$, $\pm$5$\sigma$, $\pm$7$\sigma$, ...; $\sigma$ = 24.9 m\jbm\ for \fa. Panel (b) shows the B configuration observations for \mef\ and D configuration observations for \fa. The \mef\ contours are $\pm$3$\sigma$, $\pm$5$\sigma$, $\pm$7$\sigma$, ...; $\sigma$ = 37.5 m\jbm\ and the \fa\ contours are the same as in panel (a).} \label{fig:fa-mef}}
\end{figure}

Figure~\ref{fig:fa-mef} shows a comparison of the \fa\ and \mef\ results from the CARMA B and D configurations. These maps show that the large scale emission from both species arises from roughly the same regions, while the small scale \mef\ emission is concentrated in a few small peaks at the edge of the large scale \fa\ emission.

\acal\ was only included in the B configuration observations, and it was not detected. Similarly to \fa, it is expected to have a more widespread distribution, since it has been detected only in single dish surveys \citep[e.g.][]{turner89}.  \acal\ is therefore likely resolved out by the small beam sizes used for these observations.

\clearpage

\subsubsection{Methanol [\mtoh]}\label{mtoh}
\begin{figure}[!ht]
\includegraphics[angle=270,scale=1.2]{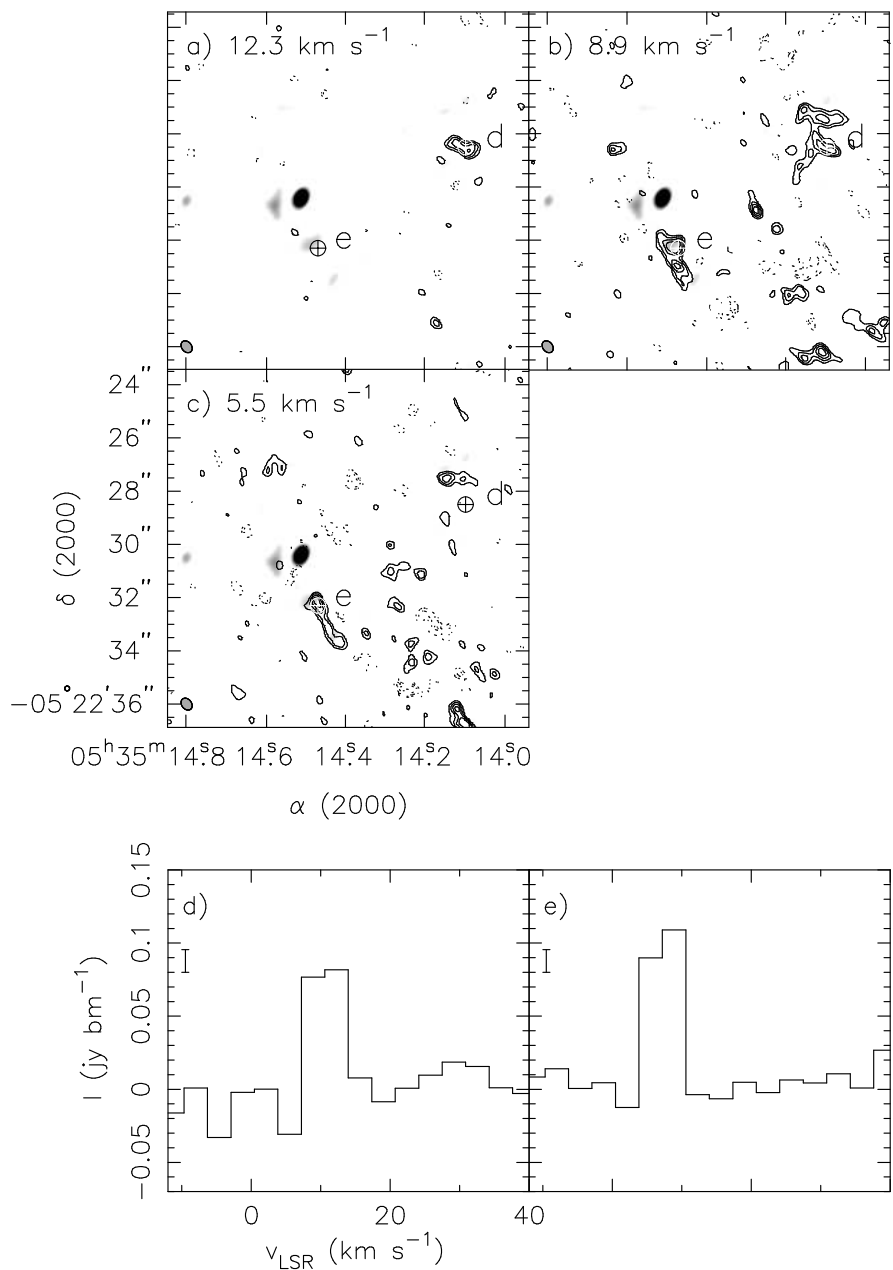}
\caption{\small{Methanol [\mtoh]  from the A configuration. Panels (a) -- (c) show maps of the \vlsr\ = 12.3, 8.9, and 5.5 \kms\ emission. Contours are $\pm3\sigma$, $\pm4\sigma$, $\pm5\sigma$,... ($\sigma$ = 15.5 m\jbm) and the gray scale is the $\lambda$ = 3 mm continuum. The spectra from these maps are shown in panels (d) and (e), with the letters on the map denoting the location of the spectra. The ``I'' bars denote the rms noise level.} \label{fig:mtoh}}
\end{figure}

\mtoh\ emission from the A configuration observations is shown in Figure~\ref{fig:mtoh}, with the emission from several velocities overlaid on the gray scale continuum. The emission at 12.3 \kms\ arises from the same regions as emission from \etcn\, and at a similar \vlsr. All of the emission peaks seen in these maps have little association with the continuum and may be arising from areas with a low dust density.

\clearpage

\subsection{Comparison of Complex Organic Molecular Species}
\begin{figure}
\includegraphics[angle=0,scale=0.8]{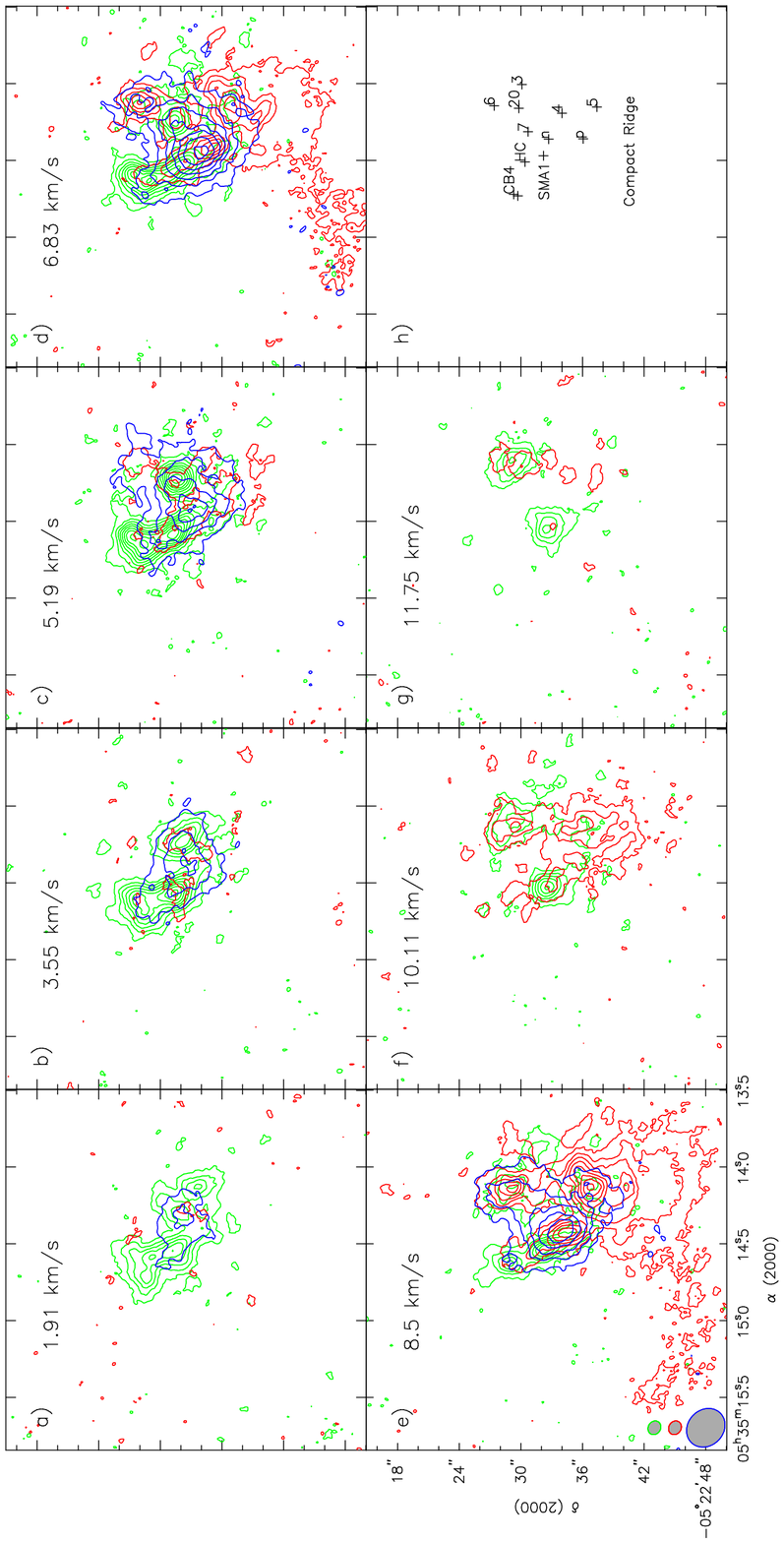}
\caption{\footnotesize{Comparison maps of \etcn, \mef, and \acetone, all created with natural weighting. The velocity for each panel is in the center top and the synthesized beams are in the lower left of panel (e). The green contours are \etcn, with levels of $\pm3\sigma$, $\pm8\sigma$, $\pm13\sigma$, ... ($\sigma$ = 12.6 m\jbm), the red contours are \mef, with the levels being the same as for \etcn, and the blue contours are \acetone, with levels of $\pm3\sigma$, $\pm6\sigma$, $\pm9\sigma$, ... ($\sigma$ = 20.6 m\jbm). Panel (h) shows notable sources in the region, see \citet{friedel11} for a more detailed description of these sources (those with number labels are IRc sources).} \label{fig:comb-vel}}
\end{figure}

A comparison between the complex molecules observed is a useful method for interpretation of the results in terms of the implications for chemical models. Figure~\ref{fig:comb-vel} shows overlayed maps of \etcn, \mef, and \acetone, all created with natural weighting. The velocity for each panel is in the center top and the synthesized beams are in the lower left of panel (e). The green contours are \etcn, with levels of $\pm3\sigma$, $\pm8\sigma$, $\pm13\sigma$, ... ($\sigma$ = 12.6 m\jbm), the red contours are \mef, with the levels being the same as for \etcn, and the blue contours are \acetone, with levels of $\pm3\sigma$, $\pm6\sigma$, $\pm9\sigma$, ... ($\sigma$ = 20.6 m\jbm). Panel (h) gives the location of the most notable sources in the region; see \citet{friedel11} for a more detailed description of these sources (those with number labels are IRc sources).

These maps enable direct comparison of the distributions of these molecules at several velocities. While the \mef\ emission is narrower in velocity space, at every velocity were \mef\ is detected, the bulk of the \mef\ and \etcn\ emission is co-spatial.  Although the emission peaks for these molecules occur at different positions, the overall emission regions are not spatially distinct as was suggested by lower spatial resolution observations. The exception to this trend is the weakly emitting, and extended, \mef\ region to the south and southeast.  Interestingly, the \acetone\ emission traces the area of greatest overlap between \mef\ and \etcn\, possibly indicating a link between \acetone\ formation and the presence of these other two species.

It is apparent from these observations that the spatial distribution of complex organic molecules in the Orion-KL region is much more complicated than has previously been considered.  Although the emission peaks arising from different molecules in Orion-KL are in some cases spatially distinct, the bulk of the emission arises from areas with spatial overlap between the molecules.  Additionally, the molecular emission is not necessarily centered on either the Hot Core or Compact Ridge sources.  These observations confirm that nitrogen/oxygen ``chemical differentiation" in its strictest definition -- i.e. where these two classes of molecules display spatially-distinct emissions peaks -- is verified.  However, previous conclusions drawn about the chemistry of this region based on the assumption that these two classes of molecules arise from spatially-distinct parcels of gas are not accurate.

The distribution of several molecules (e.g. \dme\ and \mef) also show compact emission peaks near known stars, indicating that these species are either being liberated from grain surfaces or produced in the gas phase in the warmer gas around embedded stars.  Furthermore, much of the emission attributed to the Compact Ridge has been shown, through these higher spatial resolution observations, to actually arise from nearby compact sources rather than from the more extended Compact Ridge feature identified in lower-resolution images.

As was stated above, the routine assumptions made when conducting molecular observations toward the Orion-KL region are:  1. Oxygen-bearing molecules are located in the Compact Ridge, while nitrogen-bearing molecules are located in the Hot Core;  2. The emission regions are approximately spherical and $\sim$5\arcsec~ in diameter for both sources; 3. Molecules from these sources have rotational temperatures in the 150 - 200 K range. In light of the results reported here, this view of the distribution of complex organic molecules in Orion-KL is in serious need of revision, as are the molecular abundances derived from these assumptions. Given that most of the molecules studied here display emission in many compact and spatially-distinct clumps scattered throughout the Orion-KL region, it is likely that each of these compact regions has its own distinct physical environment, spatial distribution, and chemical network.  However, interactions between these regions may in fact be driving new chemistry, as is illustrated by the comparison of the acetone distribution to those of methyl formate and ethyl cyanide.  We conclude from these results that it is no longer valid to assume all of these molecules are forming under the same chemical and physical conditions, nor is it valid to impose a simplistic model of ``chemical differentiation'' for N-bearing and O-bearing molecules in this source.  Based on these results, it is clear that chemical models should be developed that include hydrodynamic simulations so that a more realistic picture of the chemistry of star-forming regions can be developed.  Likewise, new models that properly determine the corrections for complicated spatial distributions must be developed so that molecular abundances can be adjusted to account for the observed spatial distributions.  Given that this study only included a small sample of the total chemical inventory of the Orion-KL region, additional observational studies that provide high-resolution spatial images of many different molecules are also merited.

\section{Conclusion}
We have presented CARMA observations of several molecular species, including ethyl cyanide, methyl formate, acetone, and formic acid, toward Orion-KL. These observations cover over an order of magnitude in spatial resolution, yielding excellent sensitivity to both small- and large-scale structures. From these results, we conclude that the traditional view of Orion-KL -- that of a nitrogen-oxygen "chemically differentiated" region -- is not valid for interpretation of high spatial resolution observations. The emission from nitrogen bearing and oxygen bearing species arise not only from the Hot Core and Compact Ridge, but from multiple components of the cloud that include a wide range of spatial scales and physical conditions.  Most importantly, while the emission peaks for nitrogen- or oxygen-bearing species are spatially distinct, in the majority of the emission regions these molecules are co-spatial. Additionally, several of the oxygen bearing molecules (\mef, \acetone, \fa, \& \mtoh) do not have similar distributions or peak positions. Thus, the traditional view of Orion-KL and other sources that are thought to display nitrogen-oxygen ``chemical differentiation" should be revised. Further observations of other molecular species are needed to develop a more complete chemical understanding of this source.

This study also has long-term implications for the field of astrochemistry, as it calls into question the traditional approach used when searching for new molecules.  While single-dish observations are powerful, the results presented here show that such observations must be combined with interferometric studies that probe the molecular spatial distribution.  Only these combined observational techniques will provide the information required for accurate determination of molecular abundances.  Furthermore, unbiased spectral line surveys are preferable over targeted searches, as such surveys will reveal the potentially diverse chemical composition of a source rather than targeting specific types of chemistry.  The powerful combination of spectral line surveys with interferometric observations is the next step in unraveling the mysteries of complex chemistry in star-forming regions.  Such studies are rapidly becoming possible using the next generation of correlator technology that will soon become available at the CARMA observatory, as well as those being implemented at the Extended Very Large Array (EVLA) and the Atacama Large Millimeter Array (ALMA).

\acknowledgements
We would like to thank an anonymous referee for their helpful comments. This work was partially funded by NSF grant AST-0540459 and the University of Illinois. S.L.W.W. acknowledges start-up research support from Emory University. Support for CARMA construction was derived from the states of Illinois, California, and Maryland, the Gordon and Betty Moore Foundation, the Kenneth T. and Eileen L. Norris Foundation, the Associates of the California Institute of Technology, and the National Science Foundation. Ongoing CARMA development and operations are supported by the National Science Foundation under a cooperative agreement, and by the CARMA partner universities.

\end{document}